\documentclass[sn-basic]{sn-jnl}
\usepackage{graphicx}
\usepackage{multirow}
\usepackage{amsmath,amssymb,amsfonts}
\usepackage{amsthm}
\usepackage{mathrsfs}
\usepackage[title]{appendix}
\usepackage{xcolor}
\usepackage{textcomp}
\usepackage{manyfoot}
\usepackage{booktabs}
\usepackage{algorithm}
\usepackage{algorithmicx}
\usepackage{algpseudocode}
\usepackage{listings}
\graphicspath{{images/}}

\begin{document}
\title [Modeling of the Gamma Ray Burst photospheric emission: Monte Carlo simulation of the GRB prompt emission, numerical results and discussion] {Modeling of the Gamma Ray Burst photospheric emission: Monte Carlo simulation of the GRB prompt emission, numerical results and discussion}

\author[1]{\fnm{Amina} \sur{Trabelsi}}\email{trabelsi.min76@gmail.com}

\author[2]{\fnm{Mourad} \sur{Fouka}}\email{snfouka@yahoo.fr}
\equalcont{These authors contributed equally to this work.}

\author*[1]{\fnm{Saad} \sur{Ouichaoui}}\email{souichaoui@gmail.com, souichaoui@usthb.dz, ORCID: https://orcid.org/
0000-0001-6429-2901}
\equalcont{These authors contributed equally to this work.}

\author[1]{\fnm{Amel} \sur{Belhout}}\email{abelhout@usthb.dz}
\equalcont{These authors contributed equally to this work.}

\affil*[1]{\orgdiv{Laboratory of Nuclear Sciences and Radiation-Matter Interactions (SNIRM-DGRSDT), Faculty of Physics}, \orgname{University of Sciences and Technology Houari Boumediene (USTHB)}, \orgaddress{\street{EL Alia, 16111 Bab Ezzouar}, \city{Algiers}, \postcode{32}, \country{Algeria}}}

\affil[2]{\orgname{Center for Research in Astronomy, Astrophysics and Geophysics}, \orgaddress{\street{Algiers observatory, Bouzareah}, \city{Algiers}, \postcode{63}, \country{Algeria}}}

\abstract{We have carried out a detailed study of the Gamma-Ray Burst (GRB) photospheric emission model predicting a quasi-blackbody spectrum slightly broader than a Planck function. This model was suggested within the relativistic fireball dynamics for interpreting a still not well understood thermal component in the GRB prompt emission, recently observed by the GBM (Gamma-ray Burst Monitor) on board the Fermi space telescope. We propose a Monte Carlo (M C) code for elucidating the observed spectrum, the outflow dynamics and its geometry for a basic and a structured plasma jets whose parameters are implemented. The code involves a simulation part describing the photon propagation assuming an unpolarized, non-dissipative relativistic outflow and a data analysis part for exploring main photospheric emission properties such as the energy, arrival time and observed flux of the simulated seed photons and the photospheric radius. Computing the latter two observables by numerical integration, we obtained values very concordant with the M C simulated results. Fitting Band functions to the photon spectra generated by this method, we derived  best-fit values of the photon indices matching well those featuring the observed spectra for most typical GRBs, but corresponding to fit functions inconciliable with blackbody spectral shapes. Various derived results are reported, compared to previous ones and discussed. They show to be very sensitive to the structure of the Lorentz factor that plays a crucial role in determining the presence and strength of geometrical effects. The latter manifest themselves by large broadenings of the simulated spectra featured by multiple peak energies consistently with GRB observations. They are assumed, with multiple Compton scattering, to produce bumps pointed out at very low photon energies. The interpretation of GRB observations is further performed via a Band spectral analysis using the RMFIT software. Finally, developments of this work are put into perspective.}

\maketitle
\section{Introduction} \label{Introduction} 
Despite more than five decades of extensive studies of GRBs, many questions still remain open regarding the radiation mechanisms behind these extremely intense and bright cosmological events and the complete interpretation of the relevant observations made by space telescopes. Especially over the last fifteen years, a considerable corpus of new observational data has been reported by the Fermi mission \citep{Abdo}, thanks to the large photon energy range swept by the onboard detection instruments. Indeed, this still active device launched in $2009$ hosts two broad field instruments : the GBM covering the energy range of $8$ keV $\leq E_{\gamma} \leq 40$ MeV \citep{Meegan} and the Large Area Telescope (LAT) operating in the range extending from $202$ MeV up to more than $300$ GeV \citep{Atwood}. In particular, recent observations of the GRB prompt emission made by the GBM point to a significant, still not well elucidated thermal component regarding both its origin and the underlying radiation mechanism.

On an other hand, considerable efforts have been made earlier devoted to the interpretation of GRB light curves and emission spectra proposing theoretical models \citep{Rees94, Piran, Meszaros00}, performing comprehensive spectral analyses of GRB observational data \citep{Daigne, Fouka11} detected by various instruments onboard space telescopes and/or elaborating computational analytical methods for high energy astrophysical applications \citep{Fouka09, FFouka11, FFFouka11, Fouka13, Fouka14}. However, the theoretical models based on the synchrotron emission model generally suffer shortcomings and fail to fully account for recently observed data by the Fermi-GBM. Thus, e.g., the optically thin ($\tau \ll 1$) synchrotron emission caused by internal shocks \citep{Rees94} accounts well for the non-thermal part of the observed GRB prompt emission spectrum described by a Band function \citep{Band}, $dN/dE \propto E^{\alpha}$, with typical low-energy photon index, ($\alpha \approx -1$) \citep{Preece00, Nava, ZhangB11}. But this model was found to be incompatible with the hard low energy slopes of many observed GRBs \citep{Crider, Preece98, Kaneko, Goldstein}. In addition, it fails to explain the narrow spectral width of a large fraction of GRBs \citep{Axelsson15, Yu} and the narrow distribution at a few hundred keV of their observed peak energies. 

This made the photospheric emission the most promising scenario for interpreting the observational data in the GRB prompt emission phase \citep{Thompson, Meszaros00, Rees05, Peer11, Toma, Fan, Lazzati13, Lundman, Ruffini, Deng, Begue, Gao, Peer15, Ryde17, Acuner, Hou, Meng2019, Li}. This emission process is a consequence of the relativistic fireball model where the outflow is optically thick ($\tau \gg 1$) at its base \citep{Piran}, then it becomes transparent during the fireball expansion with the trapped photons being emitted at the photosphere. The latter, related to the optical depth of the outflow, is the area (more precisely the volume) at which the photons can finally escape the plasma jet medium (see next section \ref{theory}). Furthermore, the photospheric emission model provides clarifications to the clustering of the sub-MeV peak and to the observed high efficiency prompt emission. 

Some GRB observations made by the Fermi mission, like the GRB $090902B$ \citep{Abdo, Ryde10, ZhangB11}, the GRB $100724$B \citep{Guiriec11}, the GRB $110721A$ \citep{Axelsson12}, the GRB $100507$ \citep{Ghirlanda}, the GRB $101219B$ \citep{Larsson} and the short GRB $120323A$ \citep{Guiriec13}, suggest the presence of a significant thermal component explained by the emission at the photosphere region which has a blackbody or quasi-blackbody spectrum. It has been found that most of the GRB spectra are too narrow to be explained by synchrotron emission from the electron distribution and significantly broader than a simple blackbody (Planck) spectrum. \cite{Ryde10} have found that the time-resolved spectrum of the GRB $090902B$ provides observational evidence both for a dominating photospheric emission and gradual broadening of the spectrum emitted at the photosphere. In order to clarify the spectral broadening, mainly two suggestions have been made. The first idea assumed significant photon emission below the photosphere, where the energy dissipation can modify the energy spectrum via photon scattering on the heated electrons or by possible synchrotron emission \citep{Thompson, Spruit, Rees05, Giannios, Lazzati09, ZhangYan11, Beloborodov, Vurm, Levinson}. The geometrical effects are thought as being the second issue to explain the broadening of the observed photon spectrum, which is due to the fact that the observer receives photons simultaneously emitted from different regions of the outflow with different observed temperatures because of the angle-dependence of the Doppler boost \citep{Lundman}. Then, the observed resulting spectrum is a superposition of fluxes and blackbodies at different temperatures and appears, consequently, as being highly non thermal \citep{Peer, Ito2013, Lundman, Deng}.

\begin{figure}
  \centering
  \includegraphics[scale=0.6]{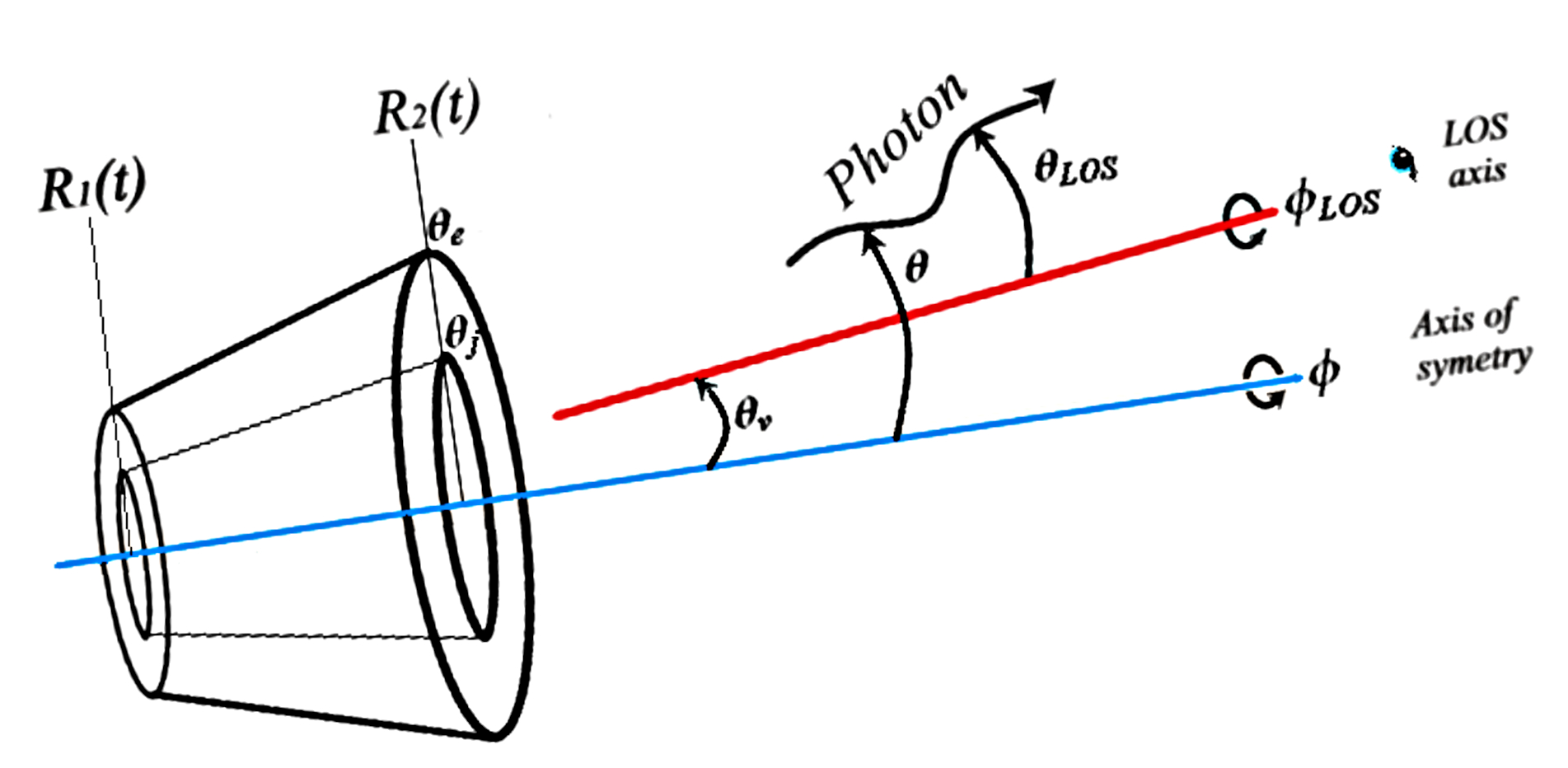}
  \caption{Schematic geometrical view of the dynamics of the jet outflow assumed to be made of two overlapped truncated cones moving versus time, where $R_{1}(t) $ and $ R_{2}(t)$ are time-dependant inner and outer boundaries counted from the origin of the jet's axis of symmetry represented by the blue line. The smaller internal cone is defined by an opening angle, $ \theta_{j} $, relative to the latter axis, while the envelope angle, $ \theta_{e} $, separates the outer jet and the envelope. The red line depicts the line of sight (LOS) direction at viewing angle, $ \theta_{v} $, relative to the jet's axis. The angle $ \theta $ locates the photon emission direction relative to the jet's axis, while $\theta_{LOS} \neq \theta + \theta_{v} $ locates the photon direction relative to the LOS direction and is obtained as the scalar product of the photon momentum vector and the LOS direction.} 
\label{fig:geometry}
\end{figure}

In this work, we have performed a detailed study of the photospheric emission model, developing a standard M C code written in C++ language composed of two distinct parts: a simulation first part describing the photon propagation through the plasma jet, and a second part devoted to data analysis. In the former part (simulation), we adopt an approach based on the plasma jet geometry for simulating the photon propagation assuming the outflow to have a conical wedge form moving in function of time, as can be seen in figure \ref{fig:geometry}. This method is explained in further details in section \ref{code}, while this associated figure serves henceforth for illustrating the forthcoming developments and visualising the geometry of the relativistic outflow. In the latter part (data analysis), we compute the photospheric radius and the observed photon flux (spectrum) by taking into account contributions from the entire emitting volume as seen by an observer located at any viewing angle, $\theta_{v}$, between the line of site (LOS) axis and the jet's symmetry axis (see figure \ref{fig:geometry}). 

We focus in this paper on the geometrical broadening of the observed photon spectrum in the GRB prompt emission phase, on the calculation of photospheric emission properties (the photospheric radius, the energy, arrival time and observed spectrum of the simulated escaped photons), their variations with main outflow free parameters (Lorentz factor, opening angle, viewing angle, power-law index p), their comparison to Fermi-GBM observational datasets and their detailed discussions.

In order to clarify the contribution of the geometrical effects, we assume that no energy dissipation occurs below the photosphere. We set two plasma jet models principally aiming to describe both the outflow dynamics and its geometry : a simple, basic plasma jet and a structured plasma jet. By studying these two plasma jet models, we aimed to gain a more detailed understanding of the likely various interplaying photon emission mechanisms at work during the GRB prompt emission phase and their dependence on the plasma jet characteristics. In the former model, we assume a relativistic spherical wind moving with constant value of the Lorentz factor after the saturation phase with a characteristic opening angle, $\theta_{j}$ $(\theta_{j} \Gamma \sim$ a $few$), (see figure \ref{fig:geometry}). In this model, the characteristic parameters of the jet such as the total luminosity, $L$, the isotropic mass ejection rate, $\overset{.}{M}$, and the baryon loading, $\eta$, are supposed to be constant but the comoving temperature of the outflow, $ T^{'}$, and the comoving electron number density, $n_{e}^{'}$, depend on the radial distance, $r$, from the outflow centre. In the structured plasma jet model, the relativistic outflow is collimated due to shocks with the surrounding gas during its expansion \citep{Aloy}. The emission comes from two regions of the outflow : the inner jet region where the Lorentz factor is approximately constant ($ \Gamma = const$) corresponding to the basic plasma jet model, and the outer jet region where $\Gamma$ is a decreasing power-law function of the angle relative to the jet's axis of symmetry (see figure \ref{fig:gammaProfile}).
Here, we adopt the same parameterization as in Refs. \citep{Zhang03, Lundman, Meng2019} for the bulk Lorentz factor angular profile and all the intrinsic parameters of the jet like the isotropic mass ejection rate, $\overset{.}{M}$, the baryon loading, $\eta$, the comoving outflow temperature, $ T^{'}$, the comoving electron number density, $n_{e}^{'}$, and the comoving photon number density, $n_{ph}^{'}$.

In the following, the theory of photospheric emission is first outlined in section \ref{theory} for describing the considered two plasma jet models, deriving required expressions to be used in each case and enlighting the subsequent code description in Section \ref{code}, where the M C simulation and data analysis parts of the code, and the numerical integration method used are presented. Then we expose and discuss in details, in section \ref{results}, the main derived results for the photospheric emission properties, while in section \ref{interpretation} we proceed to further interpretation of selected typical examples of observational data for some GRBs via fitting photospheric emission model functions to the observed photon spectra. Finally, we provide a summary and a conclusion in section \ref{conclusion} with drawing up research perspectives. Besides, a synoptic diagram of our M C simulation and numerical integration codes is presented in the Appendix.

\section{Theory of the GRB photospheric emission} \label{theory} 
\subsection{The optical depth and mean free path}
A useful concept describing the specific transparency level of a medium is that of the optical depth. In opaque environments dominated by electron scattering, the optical depth for a photon travelling along a path connecting two points $P_{1}$ and $P_{2}$ is defined by
\begin{equation}
\tau_{ray} \equiv \int_{P_{1}}^{P_{2}} n_{e} \sigma ds,
\end{equation}
where $n_{e}$ is the electron number density, $\sigma$ the photon-electron scattering cross section and $ds$ a length element along the path \citep{Rybicki}.
We assume the scattering cross section to be energy-independent and given by the Thomson cross section, $\sigma_{T}$, as long as the comoving photon energy, $ \epsilon^{'} $, is much lower than the electron mass energy, i.e., $\epsilon^{'} \ll m_{e} c^{2}$ \citep{Abramowicz}. In a more realistic situation, this assumption should be replaced as discussed by \cite{Paczynski90}.
The average optical depth travelled by a photon between two scattering events is equal to unity, i.e.,
\begin{equation}
<\tau_{ray}> \equiv \int_{0}^{\infty} \tau_{ray}  e^{- \tau_{ray}} d \tau_{ray} = 1,
\end{equation}
where $e^{- \tau_{ray}}$ is the probability for a photon to propagate an optical depth, $ \tau_{ray}$, before interacting with an electron.
A medium with large optical depth, $\tau_{ray} \gg 1$, is said to be \textit{optically thick} or \textit{opaque}: the photon cannot traverse the entire medium without being scattered (or absorbed), while when $  \tau_{ray} \ll 1$, the medium is said to be \textit{optically thin} or \textit{transparent}. 
If we consider a moving relativistic plasma with relative velocity, $ \beta = v/c = \sqrt{1 - \Gamma^{-2}} $, at an angle, $\theta_{LOS}$, relative to the line of sight (see figure \ref{fig:geometry}), the electrons occupying the space between two points, $P_{1}$ and $P_{2}$, move during the photon propagation; this motion has to be considered in the optical depth expression, which becomes	
\begin{equation} \label{optical depth}
\tau_{ray} \equiv \int_{P_{1}}^{P_{2}} \dfrac{n_{e}^{'} \sigma_{T}}{D} ds,
\end{equation}
where  $n_{e}^{'}$ is the comoving electron number density and $ D = [\Gamma (1 - \beta  \cos \theta_{LOS})]^{-1} $ the Doppler boost.
The mean free path defines as the average distance travelled by a photon through the plasma without being scattered, or the mean distance covered by the photon between two successive scatterings. It is defined in the plasma comoving frame by $l^{'} = 1/n_{e}^{'} \sigma_{T} $, and by $ l = l^{'} D $ in the observer frame.
\subsection{The photospheric emission} 
The photosphere of the outflow can be defined as the zone (a volume, see figure \ref{fig:geometry}) of the jet from where the plasma becomes transparent and the photons are free to escape from the outflow; for a specific observer, it refers to an optical depth equal to unity, i.e., $\tau_{ray} = 1$.
The location of the photosphere depends on the ratio of the baryon load, $\eta$, to a critical limiting value, $ \eta_{c} $, given by
\begin{equation}
\eta_{c} = \left(\dfrac{\sigma_{T} L \Gamma_{0}}{4 \pi r_{0} m_{p} c^{3}} \right)^{1/4},
\end{equation}
where $\Gamma_{0}$ is the initial Lorentz factor of the ejected outflow, $L$ the total luminosity, $r_{0}$ the outflow radius at the base of the ejecta, $m_{p}$ the proton mass and $c$ the velocity of light. One can distinguish two cases :\\
(i) When $\eta > \eta_{c}$, the plasma becomes transparent during the acceleration phase before reaching the coasting radius, i.e., the fireball is still accelerating with a bulk Lorentz factor, $\Gamma = r/r_{0}$  (close to  $\eta_{c}$); the emission should have a roughly blackbody shape formed by photons trapped in the outflow at radius, $r_{0}$ \citep{Paczynski86}.\\
(ii) The transparency can also be reached during the coasting phase when $\eta < \eta_{c} $. In this case, the outflow is so opaque that the radiation is trapped during the entire acceleration phase and most of the internal energy is converted into bulk motion of baryons until the Lorentz factor reaches a final constant value, $\Gamma \cong \eta$, at saturation radius, $r_{s}$ \citep{Levinson, Ghirlanda}.\\
In the following text, we adopt the second case corresponding to the photosphere located at a radius, $r_{ph} > r_{s}$; thus, the photospheric radius along the line of sight is defined as the radius where $\tau_{ray} = 1$. 
\subsection{Relativistic plasma jet models}\label{jet models} 
We consider in this work two plasma jet models. The first one is a basic relativistic steady jet within a small opening angle, $\theta_{j}$, of few degrees, optically thick at the launching region that expands in adiabatic conditions and whose energy is dominated by thermal radiation. In the second plasma jet model, we consider a similar outflow collimated by the stellar envelope that interacts with the surrounding gas, which affects the observed spectrum.
Dissipation occurring below the photosphere region may affect both the dynamics of the jet and the comoving photon spectrum. To isolate the geometrical effects, we therefore consider non-dissipative outflows cooling passively through their expansion. The plasma jet properties are geometrically expressed in a three dimensional spherical coordinate system, $(r,\theta,\Phi)$, with the polar axis aligned with the jet symmetry axis.

\subsubsection*{Basic jet model}
We consider a spherically symmetric outflow with constant total luminosity, $L$, isotropic equivalent mass ejection rate, $\overset{.}{M} = dM/dt$, and initial volume of radius, $r_{0}$. The outflow saturates and coasts with a constant value of the Lorentz factor equal to the dimensionless entropy of the outflow, $\Gamma \approx \eta$, at a so-called saturation radius given by
\begin{equation}
r_{s} \equiv \Gamma r_{0}.
\end{equation}
Above the saturation radius, the comoving temperature of the outflow defines as
\begin{equation} \label{simple plasma temperature}
T^{'} (r) = T_{0} \left( \dfrac{r_{0}}{r_{s}} \right) \left( \dfrac{r_{s}}{r} \right)^{2/3}.
\end{equation}
The temperature of the outflow at the base expresses as $T_{0} = (L/4 \pi a r_{0}^{2} c)^{1/4}$, where a is the radiation constant.
In this model, the optical depth is calculated by integrating the Eq.\ref{optical depth} and performing some approximation as done by \cite{Peer}. One obtains 
\begin{equation}
\tau(r, \theta_{LOS}) \cong \dfrac{R_{dcp}}{2 \pi r} \left(\dfrac{1}{\Gamma^{2}} + \dfrac{\theta_{LOS}^{2}}{3} \right),
\end{equation}
with $ R_{dcp} = \sigma_{T} L / (4 m_{p} \Gamma \beta c^{3})$ being  the decoupling radius where most of the photons decouple from the plasma; the latter radius is calculated by integrating the optical depth from a position in the outflow to infinity along the local radial direction and setting it to unity.\\
The photospheric radius is obtained as a function of $ \theta_{LOS} $ by setting $ \tau_{ray}(r_{ph}, \theta_{LOS}) = 1$, i.e.,
\begin{equation} \label{Rph1} 
r_{ph}(\theta_{LOS}) \cong \dfrac{R_{dcp}}{2 \pi} \left(\dfrac{1}{\Gamma^{2}} + \dfrac{\theta_{LOS}^{2}}{3} \right).
\end{equation}
To calculate the observed spectrum one has to integrate the emissivity over the entire volume where the electrons are assumed to exist. We assume that the observed photon flux depends only on the last scattering positions of the observed photons. The observed photon flux is given by
\begin{align} \label{energy flux} 
& \begin{aligned}
F^{ob}_{E}(\theta_{v})=&\int\int\frac{r^{2}}{d_{L}^{2}} D^{2} \frac{d \overset{.}{n}^{'}_{sc}}{d\Omega^{'}_{v}} exp(-\tau_{ray}) \times \lbrace E\dfrac{dP}{dE}\rbrace d\Omega dr,  
 \end{aligned}
\end{align} 
where $ d_{L} $ is the luminosity distance and $ dP/dE $ the probability density distribution in the laboratory frame describing the normalized, local photon spectrum having a comoving Planck distribution with temperature, $T^{'}$ at $r \ll R_{dcp}$. The observed spectrum in this model is calculated from an optically thick, spherically symmetric wind with $ \Gamma \approx \Gamma_{0} $ when considering only the emissivity within the angular range, $ 0 \leq\theta\leq\theta_{_{j}} $. The Doppler boost calculated using the relation 
\begin{equation} \label{Doppler boost} 
D\approx 2 \dfrac{\Gamma}{(1+ \Gamma^{2}\theta^{2})},
\end{equation}
is approximately constant for angles up to $ \theta \approx1/\Gamma_{0} $. 
The quantity, $d\overset{.}n^{'}_{sc}/d\Omega^{'}_{V} $, is the number of scattered photons in the comoving frame, expressed as      
\begin{equation} \label{scattering density} 
\frac{d\overset{.} {n}^{'}_{sc}}{d\Omega^{'}_{v}} = \dfrac{\sigma_{T} c n^{'}_{e} n^{'}_{\gamma}}{4 \pi}
\end{equation}
(the primed quantities are evaluated in the local comoving frame), where $ n^{'}_{e}$ and $ n^{'}_{\gamma}$ are the comoving electron and photon number densities, given respectively by
\begin{equation} \label{electron density} 
n^{'}_{e} = \frac{1}{r^{2} m_{p} c \beta \Gamma} \frac{d \overset{.}{M}}{d\Omega}
\end{equation}
and 
\begin{equation} \label{photon density} 
n^{'}_{\gamma} = \frac{1}{r^{2} c \Gamma} \frac{d \overset{.}{N}_{\gamma}}{d\Omega}.
\end{equation}
The photon emission rate from the base of the outflow, $ r = r_{0}$, is angle-independent, i.e., $ d\overset{.}N_{\gamma}/d\Omega = \overset{.}N_{\gamma}/4\pi $, with $ \overset{.}N_{\gamma} = L / 2.7 k T_{0}$ where $k$ is the Boltzman constant. \\
Similarly, the outflow mass rate per solid angle unit is angle-independent, i.e., $ d \overset{.}{M}/{d\Omega} = L/(4\pi c^{2}\Gamma) $.
The decoupling radius can be written as 
\begin{equation}
R_{dcp} = \dfrac{1}{(1+\beta)\beta\Gamma^{2}} \dfrac{\sigma_{T}}{m_{p}c} \dfrac{d\overset{.}M}{d\Omega}.
\end{equation}
Since $ \tau_{ray} \propto r^{-1} $, we can write $\tau_{ray} = r_{ph}/r $.
The equation (\ref{energy flux}) then becomes 
\begin{align} \label{observed spectrum} 
& \begin{aligned}
F^{ob}_{E}(\theta_{v})= &\frac{R_{dcp}}{4 \pi d_{L}^{2}} \dfrac{\overset{.}N_{\gamma}}{4\pi} (1+\beta) \int \int D^{2} \frac{1}{r^{2}} exp(- \frac{r_{ph}}{r}) \times \lbrace E \frac{dP}{dE} \rbrace d\Omega dr.
 \end{aligned}
\end{align} 
For simplicity, we approximate the photon energy distribution as a delta function centered on the average observed photon energy within volume element $ dV $, $ dP/dE= \delta(E-2.7kT^{ob}) $, where $T^{ob}$ is the observed temperature and $ 2.7kT^{ob}=((2.7DkT_{0})/\Gamma) (r_{s}/r)^{2/3}$.
The delta function variable is changed to $ r $ using the relation $ \delta(E) = \vert dE/dr \vert^{-1} \delta(r) $. Then                     
\begin{equation}
E \frac{dP}{dE} = E \delta(E-2.7kT^{ob}) = \dfrac{3}{2} r \delta \left( r-r_{s} \left\lbrace   \dfrac{D}{\Gamma \epsilon}    \right\rbrace^{3/2}  \right),  
\end{equation}
where $ \epsilon \equiv E/(2.7kT_{0}) $ is the observed photon energy in units of the average photon energy at the base of the outflow. We should underline that the observed energy of escaped photons that make their last scattering between $ \theta $ and $ \theta + d\theta $ is angle-dependent, which is expressed as $E(\theta) = 2.7kT^{ob} \left(\theta , r_{ph}(\theta) \right) = 2.7 D(\theta , \Gamma) kT^{'}\left(\theta , r_{ph}(\theta) \right) $.
Using this approximation and $ 1+ \beta \approx 2 $, $ r_{s} = \Gamma r_{0} $, the radial integral in equation (\ref{observed spectrum}) becomes
\begin{align}\label{Flux1}
& \begin{aligned} 
F^{ob}_{E}(\theta_{v})= &\frac{\overset{.}N_{\gamma}}{4 \pi d_{L}^{2}} \dfrac{3}{2} \dfrac{R_{dcp}}{r_{s}} \int_{0}^{\theta_{j}} D^{2} \left\lbrace  \dfrac{\Gamma \epsilon}{D}\right\rbrace^{3/2} \times   exp \left( -\dfrac{r_{ph}}{r_{s}} \left\lbrace  \dfrac{\Gamma \epsilon}{D} \right\rbrace^{3/2} \right) sin\theta d\theta. 
\end{aligned}  
\end{align}
At known $ \theta_{v} $, the integration can be solved numerically via the Trapeze method \citep{Press} once $\varepsilon$ and $ r_{ph} $ are calculated.
\subsubsection*{Structured plasma model}\label{structured plasma model} 
\begin{figure}
\centering
\includegraphics[scale=0.8]{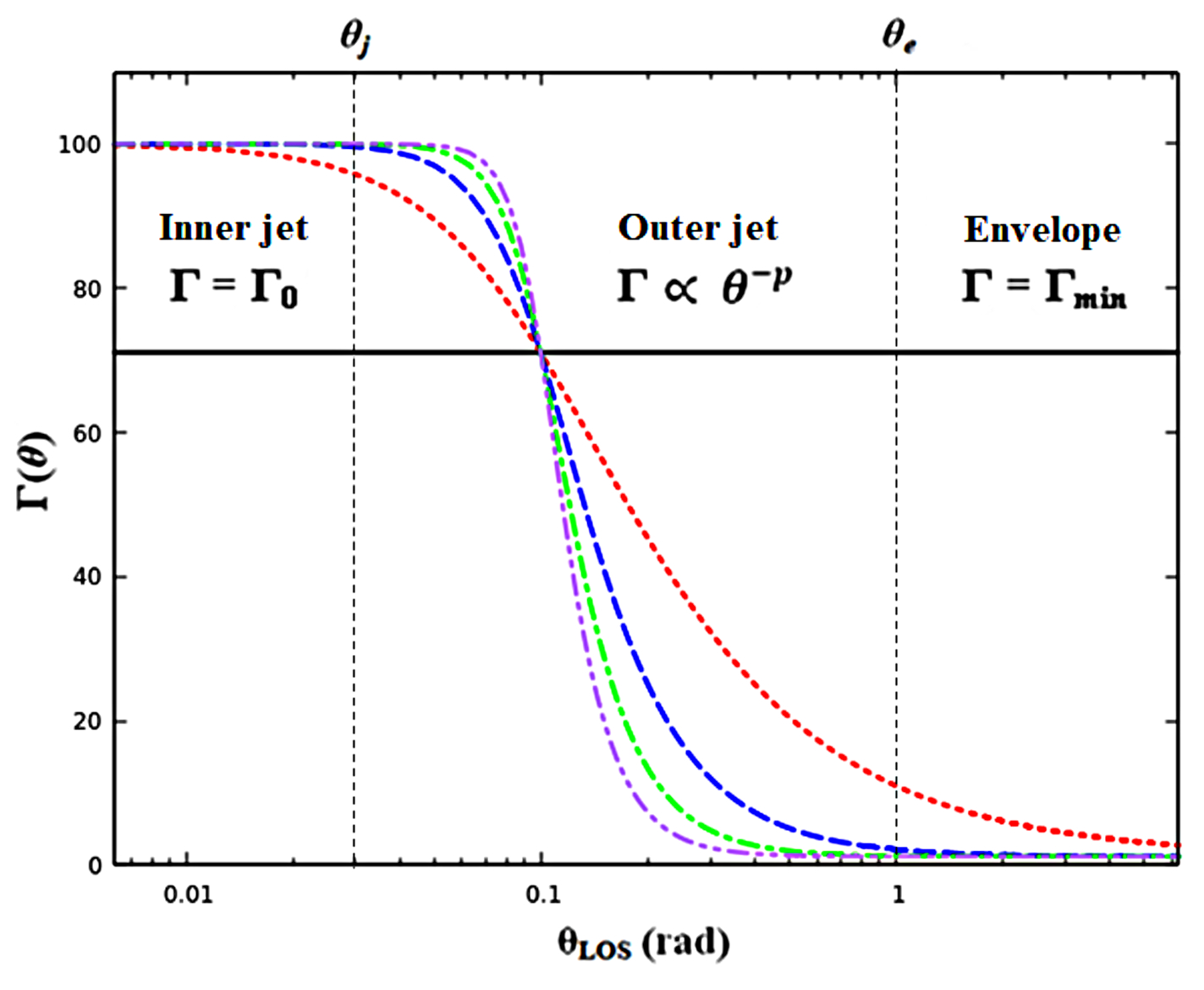}
\caption{Profile of the Lorentz factor drawn by Eq.\ref{gammatheta} for $ \Gamma_{0} = 100 $, $ \theta_{j} =1/\Gamma_{0} $ and for different values of index $ p = ( 0, 1, 2, 3, 4 )$ leading respectively to black solid, red dotted, blue dashed, green dashed dotted and purple dashed double dotted lines. The Lorentz factor is approximately constant, $ \Gamma = \Gamma_{0} $ in the inner jet region ($ \theta < \theta_{j}  $), then it decreases as a power-law with index $p$, $ \Gamma \propto \theta^{-p} $ in the outer jet region ($ \theta_{j} <\theta < \theta_{e} $). In the envelope ($ \theta > \theta_{e} $), the Lorentz factor is approximately constant with a value of $ \Gamma_{min} = 1.2 $.}
\label{fig:gammaProfile}
\end{figure}

As already stated, hydrodynamic simulations of the GRB jets after the launching phase show that the jet experiences a recollimation during its propagation through the stellar envelope, pushing material towards the sides and forming a hot cocoon that interacts with the surrounding gas. Because of the interactions with the stellar envelope, the jet edge carries more baryons than the core. This indicates that the outflow is in fact formed of two distinct regions, the inner region and a shorn layer. To study the radiative transfer in such structured plasma, a development of an angle-dependence of its properties is strongly required. Here, we adopt similar angular profiles and parameterization as extracted by \cite{Zhang03} and \cite{Lundman}. We consider an optically thick plasma in interaction with the surrounding gas having an angle-dependant baryon loading, $d \overset{.}{M} / d \Omega =d \overset{.}{M} (\theta) / d \Omega$, i.e., the outflow mass rate per solid angle, with $\theta$ being the angle relative to the jet axis of symmetry. The saturated Lorentz factor in the coasting phase is equal to the dimensionless entropy of the outflow (i.e., $\Gamma (\theta) = \eta (\theta)$) given by
\begin{equation}
\eta (\theta) = \dfrac{dL(\theta) / d \Omega}{c^{2} d \overset{.}{M} (\theta) / d \Omega},
\end{equation}
where $dL(\theta) / d \Omega$ is the jet luminosity per solid angle. Thus, for simplicity, the luminosity of the central engine is considered to be angle-independent, that is, $dL/d \Omega = L/4 \pi$, with $L$ being the total isotropic equivalent luminosity of the central engine. Therefore, the angle dependence of the saturated bulk Lorentz factor is uniquely determined by that of the baryon loading and its angular profile is taken of the form inspired from \cite{Zhang03}, i.e.,
\begin{equation}\label{gammatheta} 
\Gamma (\theta) = \dfrac{\Gamma_{0} - \Gamma_{min}} {\sqrt{ \left(\frac{\theta}{\theta_{j}} \right)^{2p}  +1}} + \Gamma_{min}.
\end{equation}
In this expression, $\Gamma_{0}$ is the maximum value of the Lorentz factor, $\theta_{j}$ the jet characteristic opening angle as stated (see figure \ref{fig:geometry}), $p$ a power-law index that determines the gradient of the profile and $\Gamma_{min} =1.2 $ the lowest value of the Lorentz factor (slightly exceeding unity for numerical reasons). The quantities $\Gamma_{0}, \theta_{j}$ and $p$ are free model parameters. From this profile, depicted in figure \ref{fig:gammaProfile}, one can distinguish two angular regions: (i) the inner jet region (for $\theta < \theta_{j}$) where the bulk Lorentz factor is approximately constant, i.e., $\Gamma \approx \Gamma_{0}$, and (ii) the outer jet region  (for $ \theta_{j} < \theta < \theta_{e}$) where $\Gamma$ is a decreasing power-law function of index $p$ of the angle, that is $\Gamma \propto \theta^{-p}$. The angle $\theta_{j}$ separating the inner and outer jet regions corresponds to the following expression of the Lorentz factor, i.e., $ \Gamma (\theta_{j}) = [ \Gamma_{0} + (\sqrt{2} - 1) \Gamma_{min}]/\sqrt{2}$, while $\theta_{e} = \theta_{j} [(\sqrt{2} + 1) \Gamma_{0} / \Gamma_{min}]^{1/p}$ is the characteristic envelope angle separating the outer jet region and the envelope region, $( \theta > \theta_{e})$, where the Lorentz factor is given by $ \Gamma (\theta_{e}) = \sqrt{2} \Gamma_{min} $.\\
The electron number density and the photon number density in the local comoving frame are angle-dependant and given, respectively, by the expressions 
\begin{equation} \label{electron number density}
n_{e}^{'} (r, \theta) = \dfrac{1}{r^{2} m_{p} c \beta (\theta) \Gamma (\theta)} \dfrac{d \overset{.}{M} (\theta)}{d \Omega}
\end{equation} and
\begin{equation} \label{photon number density}
n_{\gamma}^{'} (r, \theta) = \dfrac{1}{r^{2} c \Gamma (\theta)} \dfrac{d \overset{.}{N_{\gamma}} (\theta)}{d \Omega}.
\end{equation}
Since the saturation radius given by $ r_{s} (\theta) = \Gamma(\theta) r_{0} $ is angle-dependant, the comoving temperature of the jet at angle $\theta$ and radius $r > r_{s} (\theta)$ expresses as                  
\begin{equation} \label{structured plasma temperature}
T^{'} (r, \theta) = T_{0} \left(\dfrac{r_{0}}{r_{s} (\theta)}\right) \left(\dfrac{r_{s} (\theta)}{r}\right)^{2/3}.
\end{equation}
The angle-dependant outflow mass rate per solid angle is expressed by $ d\overset{.}M(\theta)/d\Omega = L/4 \pi c^{2} \Gamma(\theta) $. However, the photon emission rate is simplified as being angle-independent, i.e., $ d\overset{.}N_{\gamma}(\theta) = \overset{.}N_{\gamma}/4\pi $. Eq.\ref{electron number density} of the electron number density allows the decoupling radius to be written in the form 
\begin{equation} \label{decoupling radius} 
R_{dcp}(\theta) = \dfrac{L \sigma_{T}}{4 \pi m_{p} c^{3}} \dfrac{1}{(1+\beta(\theta)) \beta(\theta) \Gamma(\theta)^{3}}. 
\end{equation}	
We obtain the following expression of the optical depth, i.e.,
\begin{align}
& \begin{aligned}
\tau_{ray} = &\dfrac{\sigma_{T}}{m_{p} c r \sin \theta_{LOS}} \int_{0}^{\theta_{LOS}} \dfrac{(1 - \beta(\overset{\sim}{\theta}_{LOS})\cos \overset{\sim}{\theta}_{LOS})}{\beta(\overset{\sim}{\theta}_{LOS})} \dfrac{d \overset{.}{M}(\overset{\sim}{\theta}_{LOS})}{d \Omega} d \overset{\sim}{\theta}_{LOS},
 \end{aligned}
\end{align}
from which we deduce the expression of the photospheric radius, i.e.,
\begin{align} \label{Rph2} 
& \begin{aligned}
r_{ph} = &\dfrac{\sigma_{T}}{m_{p} c \sin \theta_{LOS}} \int_{0}^{\theta_{LOS}} \dfrac{(1 - \beta(\overset{\sim}{\theta}_{LOS})\cos \overset{\sim}{\theta}_{LOS})}{\beta(\overset{\sim}{\theta}_{LOS})} \dfrac{d \overset{.}{M}(\overset{\sim}{\theta}_{LOS})}{d \Omega} d \overset{\sim}{\theta}_{LOS}.
\end{aligned}
\end{align}
In this model, the emissivity is considered from the optically thick plasma interacting with the surrounding gas, so that the jet is structured in two angular regions, the inner region at $ 0\leq \theta\leq \theta_{j} $ and the outer region at $ \theta_{j}\leq \theta\leq \theta_{e} $. Thus, one can obtain the observed flux from $ 0\leq \theta\leq \theta_{e} $, considering the angular profile of the Lorentz factor as given by the equation (\ref{gammatheta}). Using the angle dependences mentioned above, the expression of the observed flux can be written as
\begin{align}
& \begin{aligned}
F^{ob}_{E}(\theta_{v})= &\dfrac{1}{4 \pi d_{L}^{2}} \frac{\overset{.}N_{\gamma}}{4 \pi} \int \int (1+\beta)D^{2} \times \dfrac{R_{dcp}}{r^{2}} exp  \left( -\dfrac{r_{ph}}{r}\right) \left\lbrace E \dfrac{dP}{dE} \right\rbrace d\Omega dr. 
\end{aligned}  
\end{align}
To simplify the calculation, we use the approximations given in the previous section, and we get 
\begin{align}
& \begin{aligned}
F^{ob}_{E}(\theta_{v})=& \frac{\overset{.}N_{\gamma}}{4 \pi d_{L}^{2}} \dfrac{3}{2} \int_{0}^{\theta_{e}} D^{2} \dfrac{R_{dcp}}{r_{s}} \left\lbrace  \dfrac{\Gamma \epsilon}{D}\right\rbrace^{3/2} \times exp \left( -\dfrac{r_{ph}}{r_{s}} \left\lbrace  \dfrac{\Gamma \epsilon}{D} \right\rbrace^{3/2} \right) sin\theta d\theta. 
\end{aligned}  
\end{align}
The Lorentz factor, $ \Gamma $, the Doppler boost, $ D $, the decoupling radius, $ R_{dcp} $ and $ r_{ph} $ are well approximated by power-laws of angle $ \theta $ as has been shown by \cite{Lundman}, i.e.,
\begin{center}
$\Gamma(\theta) = \Gamma_{0} \left( \dfrac{\theta_{j}}{\theta}\right)^{p}  $,\\
$ D(\theta) = 2\Gamma_{0} \left( \dfrac{\theta_{j}}{\theta} \right)^{p}$,\\
$ R_{dcp}(\theta) = \dfrac{R_{\ast}}{\Gamma^{3}_{0}} \left( \dfrac{\theta}{\theta_{j}}\right)^{3p} $,\\
$ r_{ph}(\theta) = \dfrac{R_{\ast}}{(3p+1)\Gamma^{3}_{0}} \left( \dfrac{\theta}{\theta_{j}}\right)^{3p} $,
\end{center}
with $ R_{\ast} \equiv L \sigma_{T}/8 \pi m_{p} c^{3} $. 
Then, the observed flux can be written as
\begin{align}
&\begin{aligned}
F^{ob}_{E} = & \dfrac{\overset{.}N_{\gamma}}{4 \pi d_{L}^{2}} \dfrac{3}{2} \dfrac{4}{\left(  \Gamma_{0} \theta^{p}_{j}\right)^{2} } \dfrac{R_{\ast}}{r_{0}} \left(\dfrac{\epsilon}{2}\right)^{3/2} \int_{\theta_{s}}^{\theta_{D}} \theta^{2p+1} \times exp \left( - \dfrac{1}{3p+1} \dfrac{1}{(\Gamma_{0} \theta^{p}_{j})^{4}}   \frac{R_{\ast}}{r_{0}} \left( \frac{\epsilon}{2}\right)^{3/2} \theta^{4p}\right) d\theta, 
\end{aligned} 
\end{align}
which leads to 
\begin{align}
&\begin{aligned}\label{flux2} 
F^{ob}_{E} = & \dfrac{\overset{.}N_{\gamma}}{4 \pi d_{L}^{2}} \dfrac{3}{2} \dfrac{\left( 3p +1 \right)^{1/2 \left( 1/p +1 \right)} } {p} \left( \Gamma_{0} \theta^{p}_{j} \right)^{2/p} \left( \dfrac{R_{\ast}}{r_{0}} \right)^{1/2 (1-1/p)} \times \left( \dfrac{\epsilon}{2} \right)^{3/4 (1-1/p)} \\
 &\int_{x_{min}}^{x_{max}} x^{1/2 (1/p -1)} exp (-x) dx,
\end{aligned} 
\end{align}
where 
\begin{center}
 $x \equiv (3p +1)^{-1} (\Gamma_{0} \theta^{p}_{j})^{-4} (R_{\ast}/r_{0}) (\epsilon/2)^{3/2} \theta^{4p}$,\\
 $x_{min} = (3p +1)^{-1} (R_{\ast}/r_{0}) (\epsilon/2)^{3/2} (\Gamma_{0} \theta^{p}_{j})^{4/p-1}$,\\
 $x_{max} = (3p +1)^{-1} (R_{\ast}/r_{0}) (\epsilon/2)^{3/2} ((1+\sqrt{2})/ \Gamma_{min})^{4}$  .
\end{center}
Setting the upper limit to infinity (which is a good approximation for $ x_{max} \gg 1 $), the latter integral can be written as \\
$\int_{x_{min}}^{\infty} x^{1/2 (1/p -1)} exp (-x) dx = \Gamma_{inc} \left\lbrace \dfrac{1}{2} \left( \dfrac{1}{p} +1 \right), x_{min} \right\rbrace $, where  $ \Gamma_{inc} \left\lbrace \dfrac{1}{2} \left( \dfrac{1}{p} +1 \right), x_{min} \right\rbrace $ is the incomplete gamma function calculated via the Boost library. The photospheric radius and the observed flux are calculated by numerical integration using the Trapeze method \citep{Press}.

\section{Numerical simulation and code descriprion} \label{code} 
In order to compute the light curves and energy spectra for the GRB photospheric emission from relativistic, collimated and non-magnetized jets characterized by the two plasma jet models described above, we have developed a standard numerical M C simulation code consisting in two parts (see Appendix): a simulation part that trucks the photon propagation inside the outflow and a data analysis part. After simulating a sufficient number of photons, we collect the escaped ones and introduce them into the latter part. Below in this section, we expose in details the procedures used in these two parts of the code.
\subsection{Simulation part} 
In this part, we characterize a photon by its time, position and three-momentum vector from which the photon direction and its energy can be deduced. The photon position and its direction are expressed in a three dimensional spherical coordinates system, (r,$\theta$,$\phi$). Based on the M C simulation method, we produce the initial photon by randomly generating its position and its three-momentum vector from a uniform distribution. Its initial comoving energy is drawn from a Planck distribution with the comoving plasma temperature at the photon position expressed by Eqs.\ref{simple plasma temperature} and \ref{structured plasma temperature} respectively in the cases of the basic and structured plasmas. The seed photons are then initially generated following a blackbody law and injected into the plasma with $\tau_{ray} = 20$ to ensure a low probability of $\exp (-20) \approx 2.1 \times10^{-9}$ for a photon to escape without being scattered. 
         
The scattered electron is expressed via its randomly generated three-momentum vector (direction). Its energy is derived from a relativistic Maxwell distribution with the plasma temperature calculated at the photon position by Eqs.\ref{simple plasma temperature} and \ref{structured plasma temperature} respectively for a basic and a structured plasmas. 

The photon travels along a straight line between two successive scatterings in the lab frame but its energy and direction upon scattering are most easily obtained in the electron rest frame, which leads us to choose the former frame for the photon propagation and the latter frame for its scattering. Hence, in the scattering process the code performs a Lorentz transformation of the photon properties from the lab frame to the electron rest frame via the local comoving frame. 
Starting from the properties of the initial photon in the lab frame, an iterative process of its propagation is made. Primarily, one verifies if the photon is located within the plasma by checking its position via the geometrical simulation conditions (see next subsection). Otherwise, it is supposed to be somewhere else out of the jet. 
The photon propagates a certain distance in the lab frame until it interacts with an electron via Compton scattering. Then, the scattering position is calculated and taken as the new photon position. The full Klein-Nishina interaction cross section is used for calculating the energy and the three-momentum vector of the scattered photon in the electron rest frame. Finally, the photon properties are transformed back into the lab frame, and the process is repeated until the photon escapes from the plasma. 
									
After the simulation of a sufficient number of photons ($10^{6}$), we proceed to the collection of those having escaped from the plasma, i.e., having diffused since the last scattering event, each photon being featured by its own properties, i.e., its position at last scattering, three-momentum vector, direction and energy. This photon collection is thereafter considered in the analysis part of the code (see section \Ref{analysis} and Appendix).

\subsubsection*{Geometrical simulation}\label{geometricalSim} 
The common simulation of the photon propagation in previous works (see, e.g., \cite{Peer}, \cite{Lundman} and \cite{Begue13}) is performed by comparing the optical depth, $\tau_{ray}$, travelled by the photon before it escapes the plasma and the optical depth, $\Delta \tau$, representing the distance between two successive scattering events in the direction of propagation derived from a logarithmic distribution. If $\Delta \tau \geq \tau_{ray}$, the photon  escapes the outflow; otherwise it remains trapped in the latter and continues its propagation. This method requires the calculation of $\tau_{ray}$ that depends on the angle $\theta_{LOS}$, which is a crucial parameter in the final data analysis part of the code.

As stated, we adopt in this work an approach based on the jet geometry for simulating the photon propagation. As shown in figure \ref{fig:geometry} presented in section \ref{Introduction}, we assume that the outflow has a conical wedge form moving in function of time. For a more realistic picture, it is a truncated cone geometrically limited by the jet opening angle, $\theta_{j}$, and the time-dependant inner and outer boundaries, $R_{1}(t) = r_{0} + vt$ and $R_{2}(t) = R_{1}(t) + Th$ where $v$ and $Th$ denote respectively the plasma velocity and the outflow thickness, as shown in this figure. For each photon position characterized by $r$ and $\theta$, we verify two geometrical conditions:\\
(i) if $R_{1}(t) < r < R_{2}(t)$ and $ \theta < \theta_{j}$ (for the basic plasma jet model) or $\theta < \theta_{e}$ (for the structured plasma jet model), the photon is assumed to be trapped inside the outflow and the scattering is assumed to occur,\\
(ii) otherwise, the scattering does not take place and the photon is considered as being decoupled. Then, the loop is repeated until the photon comes out from the outflow.\\
Here, we simulate the propagation of a photon thought an outflow with the quantities $L, r_{0}, \theta_{j}, \Gamma_{0}$ and the thickness, $Th$, being free profile parameters, as described in the basic plasma jet model. Besides, we set these same quantities together with the power-law index, $p$, as free profile parameters in the structured plasma jet model. At the end of the simulation, we collect the emitted photons escaped from the jet at the photosphere region and inject all the data in the analysis part of the M C code. Notice that one can actually introduce any plasma jet model in the simulation part of our M C code (see the Appendix).
\subsection{Analysis part} \label{analysis}
In this part, we recover the collection of the escaped photons generated by the simulation via both two jet's plasma models described above. The treatment of the properties of the escaped photons does not depend on the considered plasma jet model. We previously calculated the properties (position vector, three-vector momentum and plasma velocity vector) of each escaped photon in the lab frame and in the Cartesian coordinates system. The photon emission time, its energy in the comoving frame, the number of photon scatterings and other useful characteristic quantities are also calculated as follows in considering the photon scattering process.\\
To study the energy transfer, we calculate the Compton ratio as 
\begin{equation}
 Y = \frac{E^{'}}{E_{i}},
\end{equation} 
where  $E^{'}$ and $ E_{i} $ are respectively the emitted photon energy and its initial energy in the comoving frame. We also add a weight, $ w $, to each emitted photon, calculated as the ratio,
\begin{equation}
 w = \frac{N_{tot}}{N_{phBin}}.
\end{equation} 
of the total number of photons, $ N_{tot} $, emitted in a time interval, $ \Delta t $, to the number of photons per bin, $ N_{phBin} $ (i.e., the ratio of the number of analysed photons to the number of bins). The total number of photons in the preceding expression is calculated by the relation 
\begin{equation}
 N_{tot} = \Delta t A \textit{J},
\end{equation} 
where $ A $ is the outflow area, given by 
\begin{equation}
A= 4\pi r^{2} \sin \theta  d\theta + 2\pi \sin \theta  r dr,  
\end{equation} 

and $ \textit{J} $ denotes the emission coefficient expressed as 
\begin{equation}
\textit{J}= \left( \dfrac{2}{c^{2}}\right)  2\xi(3) \left( \dfrac{k T^{'}}{h}\right)^{3},
\end{equation} 
in term of the Riemann function, $ 2\xi(3) = 2.40411 $.
The position of the escaped photon is the place where the latter is emitted, corresponding to the photon last scattering. Therefore, one can obtain the photospheric radius by calculating the magnitude of the position vector,
\begin{equation}
r_{ph}= \sqrt{x^{2} + y^{2} + z^{2}},
\end{equation}  
whose associated unit vector in the direction of the plasma parcel on the photosphere is $ \overrightarrow{e_{r}} $. We deduce the escaped photon direction by extracting the unit vector, $ \overrightarrow{e_{_{m}}}$, associated with the three-vector momentum. From the plasma velocity vector,  $\overrightarrow{v_{p}}$, we obtain the plasma direction vector, $\overrightarrow{e_{p}}$, the plasma relative velocity, $ \beta_{p} = V_{p}/c $, and the associated Lorentz factor, $ \Gamma_{_{p}}=1/\sqrt{1-\beta_{p}^{2}} $.
We set two free parameters in this analysis : the redshift, z, and the viewing angle, $\theta_{v} $, between the LOS axis and the jet symmetry axis (already defined in section \ref{Introduction}, see also figure \ref{fig:geometry}). Then, one can draw the observer's direction with associated unit vector, $\overrightarrow{e_{obs}}$. We compute the following three distinct angles : $\theta_{r}$, $\theta_{LOS}$ and $\theta_{p}$, respectively between the latter direction (unit vector) and the position vector, the momentum vector and the plasma direction, whose associated respective unit vectors are $\overrightarrow{e_{r}}$, $\overrightarrow{e_{_{m}}}$ and $\overrightarrow{e_{p}}$. Those angles are given by the following expressions :
\begin{center}
$\theta_{r}= \arccos (\overrightarrow{e_{r}}, \overrightarrow{e_{obs}})$,\\
$\theta_{LOS}= \arccos (\overrightarrow{e_{_{m}}}, \overrightarrow{e_{obs}})$,\\
$\theta_{p}= \arccos (\overrightarrow{e_{p}}, \overrightarrow{e_{obs}})$.
\end{center}
Therefore, we can derive the Doppler boost, $ D $, the arrival time, $ t_{a}$, and the final weight, $w_{\textit{f}}$, respectively by the following expressions:	
\begin{equation}
D= \dfrac{1}{\Gamma_{p} (1 - \beta_{p}\cos \theta_{p} )},
\end{equation}
 \begin{equation}\label{ArrivalTime} 
t_{a}= \left(t_{em}-\dfrac{r_{ph}}{c} \cos \theta_{r}\right)(1 + z)(1 - \beta\cos \theta_{r})
\end{equation}
and
 \begin{equation}
 w_{\textit{f}}= w \cos \theta_{LOS} \frac{1}{4\pi d_{L}^{2}},
\end{equation}
where $t_{em}$ is the emission time obtained directly from the simulation part's data collection. 
The observed photon energy is calculated by the relation
 \begin{equation}
E_{obs}=\dfrac{D E_{\textit{cf}}}{(1 + z)},
\end{equation}
where $ E_{\textit{cf}} $ denotes the emitted photon energy in the comoving frame.
The observed flux is derived by the expression 
\begin{equation}
F_{obs}= w_{\textit{f}} D^{3} I_{_{cf}}, 
\end{equation}
where $ I_{cf} $ is the comoving photon number intensity, given by
 \begin{equation} \label{intensity} 
I_{_{cf}} = \frac{N_{tot}}{4 \pi r_{_{ph}}^{2} \Gamma_{p}} \times  \frac{1}{4 \pi} \times \frac{4}{(1+\beta_{_{p}})^{2}}.
\end{equation}
This last equation thus involves three factors 
: the comoving photon emission rate through a sphere of radius, $ r_{ph} $, its isotropic intensity and a correction factor due to only considering the outward radiation in the laboratory frame.

\section{Results and discussion}\label{results} 
In the simulation part of our code, we have assumed an unpolarized, non-dissipative dynamics with a large profile of free parameters for both two considered astrophysical plasma jet models (the basic and the structured plasma jets), setting $ L = 10^{52} erg s^{-1}$, $r_{0} = 10^{6}$ cm, $\Gamma_{0}=100$, $ 300$, $ 500$, $ \Gamma_{0} \theta_{j}= 1$, $3$, $5$, $10$ and $Th = 10^{20}$ cm. In the case of the structured plasma jet, we have added the power-law index values $ p = 1$, $2$, $4$. A number of $ 10^{6} $ photons have been simulated in each run.

In the analysis part, we have set $ \theta_{v}/\theta_{j}$ values of $0$, $0.5$, $1$, $1.5$ and $2$, $z = 2$, which leads to a luminosity distance value of $ d_{_{L}} = 4.318$ Gpc used for the spectrum normalization. The derived results (histograms) are presented below in logarithmic scale.

\subsection{Basic plasma jet model}\label{basic results} 
\begin{figure}
\centering
\includegraphics[scale=0.8]{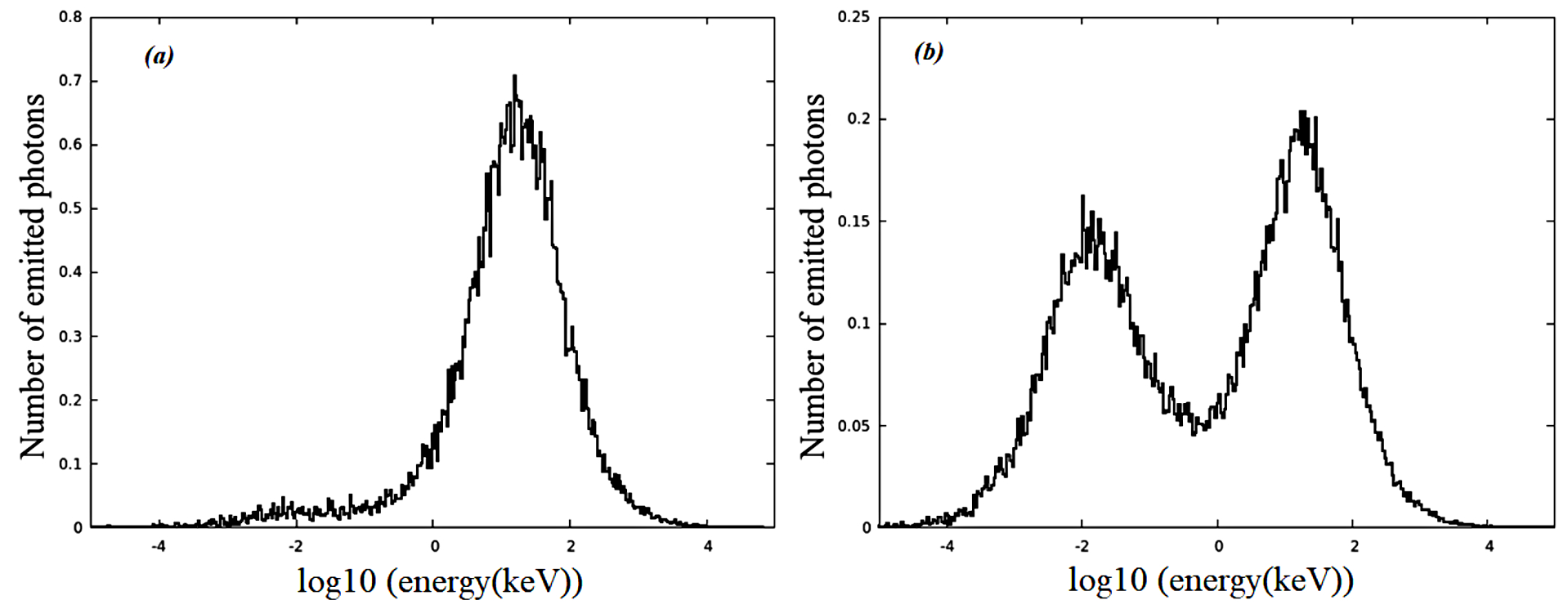}
\caption{Derived photon energy distributions. Panel (a) : energy distribution for initial Lorentz factor, $\Gamma_{0} = 500$.  Panel (b): energy distributions for $\Gamma_{0} = 100$. An opening angle $\theta_{j} =10/\Gamma_{0}$ and zero viewing angle, $\theta_{v} = 0$, are assumed in these plots.}
\label{fig:energy1}
\end{figure}

\begin{figure}
\centering
\includegraphics[scale=0.6]{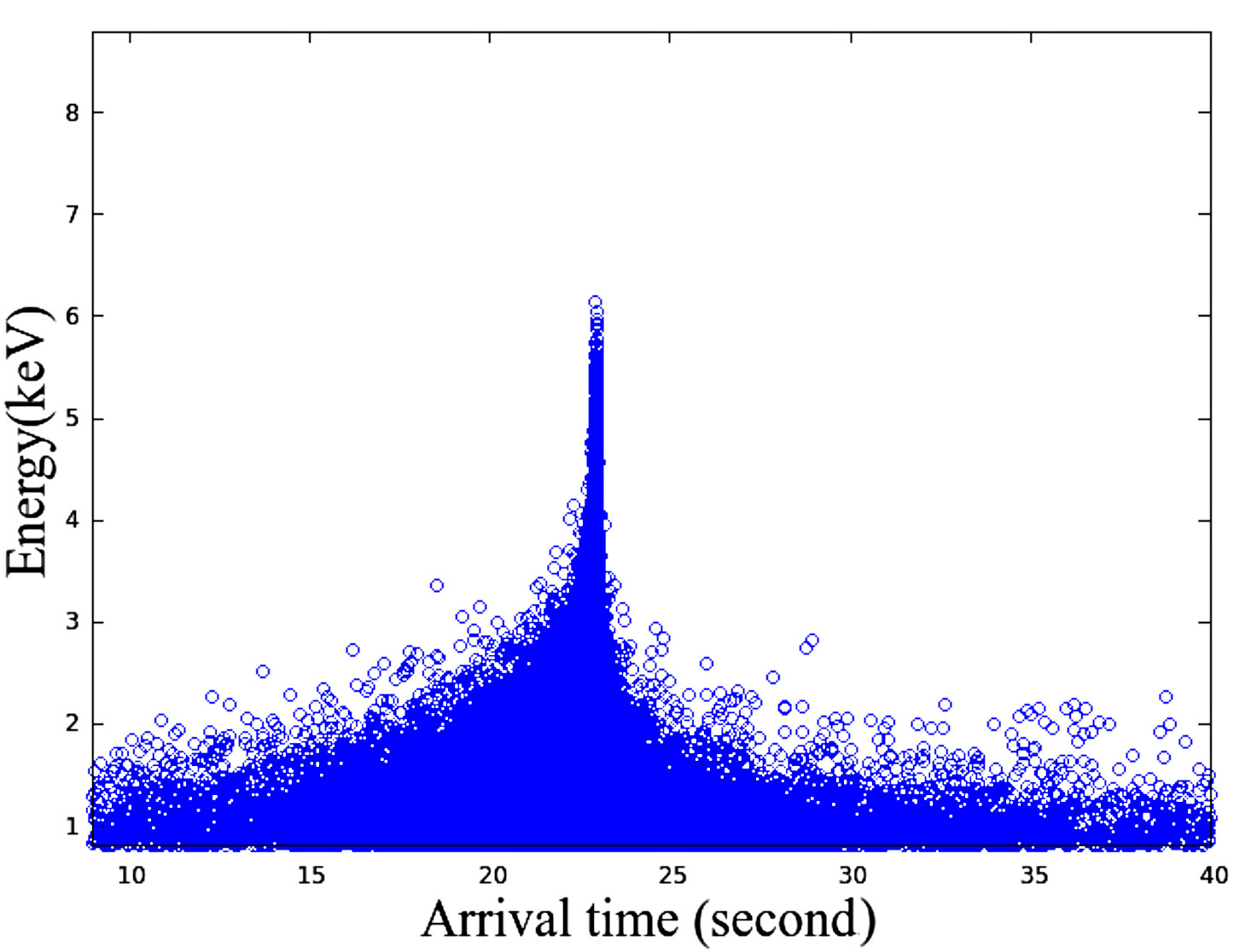}
\caption{Variation of the photon energy versus the photon arrival time for $\Gamma_{0} = 100$, $\theta_{j} =10/\Gamma_{0}$. A zero viewing angle, $\theta_{v} = 0$, is assumed in this plot.}
\label{fig:energyTime}
\end{figure}

\begin{figure}
\centering
\includegraphics[width=1.\textwidth]{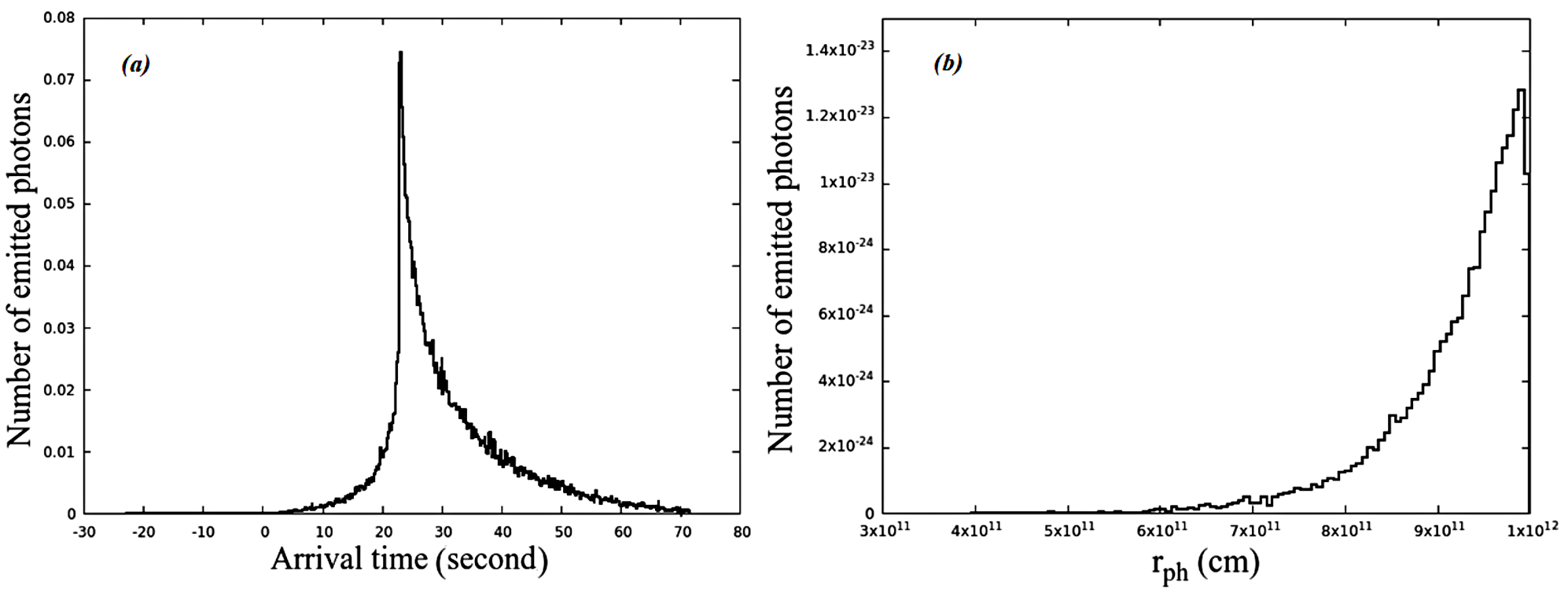}    
\caption{Observed arrival time and photospheric radius histograms (panels (a) and (b), respectively) presented for $\Gamma_{0} = 100$, $\theta_{j} =10/\Gamma_{0}$ and $\theta_{v} = 0$.}
\label{fig:TimeRph1}
\end{figure}

\begin{figure}
\centering
\includegraphics[width=1.\textwidth]{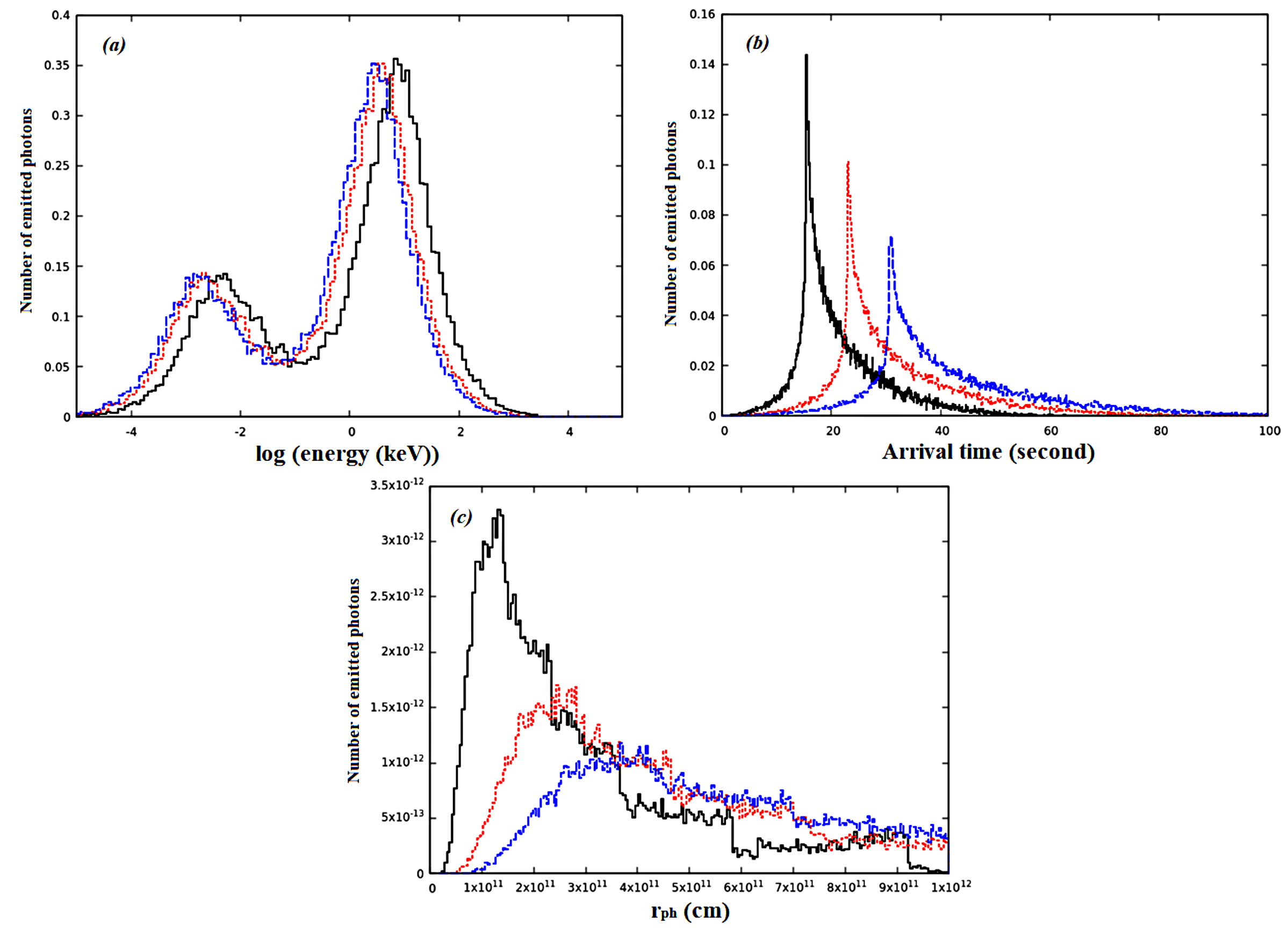}
\caption{Photon energy, arrival time and photospheric radius histograms (panels (a), (b) and (c), respectively) for three different values of the viewing angle, $\theta_{v}$. We put $\theta_{v} = few \times \theta_{j}$, where $few = 0$, $1$ and $2$, leading, respectively, to the black solid, red dotted and blue dashed spectra. In these plots, we took $\Gamma_{0} = 300$ and $\theta_{j} = 1/\Gamma_{0}$.}
\label{fig:VariationThetav1}
\end{figure}

In the case of the basic plasma jet model, we report in figures \ref{fig:energy1} up to \ref{fig:VariationThetav1} the histograms derived in the spectral analysis for the photon energy, the photon arrival time and the photospheric radius. In figures \ref{fig:energy1} to \ref{fig:TimeRph1}, we considered a narrow jet with opening angle $\theta_{j} = 10/\Gamma_{0}$,  $ \Gamma_{0} = 100 $ (except for the panel (a) of figure \ref{fig:energy1} where $ \Gamma_{0} = 500 $), that is $ \theta_{j} = 0.1 $ rad. This is a typical, well constrained value of the opening angle commonly used in GRBs modeling, corresponding to a negligible impact of the structure of the jet’s luminosity on the spectrum calculation when the viewing angle, $ \theta_{v} $, is smaller than the core angle, $ \theta_{c,L} $, referring to isotropic angular width of the jet associated with luminosity \citep{Meng2024}. A zero viewing angle is assumed in the analysis, which means that the observer is looking directly at the GRB jet. In other words, the photons from the jet are weakly Doppler-boosted and their energy is the same as the energy of the photons emitted by the electrons within the jet.

The photon energy distributions are presented in figure \ref{fig:energy1} (panels (a) and (b) where the x-axis denotes the logarithm in basis 10 (log) of the photon energy, $E_{\gamma}$, expressed in keV units). In the panel (a), the energy distribution is plotted for initial Lorentz factor of $\Gamma_{0} = 500$. As can be seen, a prominent peak is identified at log ($E_{\gamma}$) $\sim 1.22 $ corresponding to a photon energy value of around $16.59$ keV. This is very consistent with observations for the GRB $090708$ revealing a similar photospheric emission with the peak energy lying at $17.4 \pm 0.64$ keV, as tabulated in \cite{Nava}, and the GRB $090815B$ with a peak energy located at around $ 15 \pm 10.8$ keV observed by the Fermi-GBM (see the Fourth Fermi-GBM GRB Catalog in \cite{VonKielen}).
A low-energy photon emission exhibiting a peak energy of less than $50$ keV indicates that the photosphere was very hot, with a temperature reaching about $10^{9}$ K. This is considerably hotter than in the case of a star's photosphere featured by a much lower temperature value typically amounting to about $5000$ K. The high temperature of the photosphere is thought to be due to the fact that the burst was produced by the collapse of a massive star that can considerably heat up the photosphere.

The panel (b) of this figure where $\Gamma_{0} = 100$ shows a prominent, main peak at log ($E_{\gamma}$) $\sim 1.28$ corresponding to $E_{\gamma}$ $\sim 19.05$ keV and a smaller peak located at very low energy, i.e., log ($E_{\gamma}$) $\sim -1.87$ corresponding to $E_{\gamma}$ $\sim 0.013$ keV. A clear evidence for the presence of the latter peak appearing as a bump at very low $E_{\gamma}$ is also manifested in figure \ref{fig:energyTime} reporting the variation of the photon energy versus its arrival time for the same outflow free parameters ($\Gamma_{0} = 100$, $\theta_{j} =10/\Gamma_{0}$, $\theta_{v} = 0$). These values are in excellent agreement with observational data for observed GRBs : the GRB $ 170516A $ observed by the Neil Gehrels Swift Observatory\footnote{https://swift.gsfc.nasa.gov/results/batgrbcat/GRB170516A/web/GRB170516A.html} with a peak energy at $ 19.1 \pm 12.4$ keV, the GRB $ 130702A $ and the GRB $081124A$ with peak energies respectively at $ 20 $ keV and $22.8 \pm 0.7$ keV, detected by the Fermi-GBM (see the Fourth Fermi-GBM GRB Catalog in \cite{VonKielen}).

This bump emerging at very low photon energy before the main peak can be explained by invoking several scenarios : sub-photospheric energy dissipation processes (synchrotron radiation or inverse Compton scattering), thermal emission from the shock-heated material in the vicinity of the burst site, geometrical effects, multiple Compton scattering. The latter process can produce aberrations at very low energy of the photon spectrum, potentially modifying the light curve and leading to the formation of bumps or deviations in it \citep{aksenov}. This is essentially due to the fact that photons experiencing multiple Compton scatterings lose a greater cumulative amount of their energy, which causes a fluctuating low-energy tail in the observed spectrum and can distort the overall photon energy distribution. This distortion may influence the emission spectrum, eventually broadening its shape and producing unexpected bumps at extremely low energies. In this work, we have performed a geometrical simulation considering the entire volume of a non-dissipative relativistic jet using the spherical coordinates, which then excludes sub-photospheric energy dissipation effects. Finally, we think that the low-energy bump generated in our M C simulation is very likely due mainly to geometrical effects and multiple Compton scattering. This process has been previously invoked for interpreting the appearance of this bump at the extremely low energy part of the photon spectrum and pointed out in some GRB observations as reported by \cite{Lazzati09}, \cite{beloborodov2011}, \cite{aksenov} \cite{mushtukov2022}. 

\begin{figure}
\centering
\includegraphics[width=1.\textwidth]{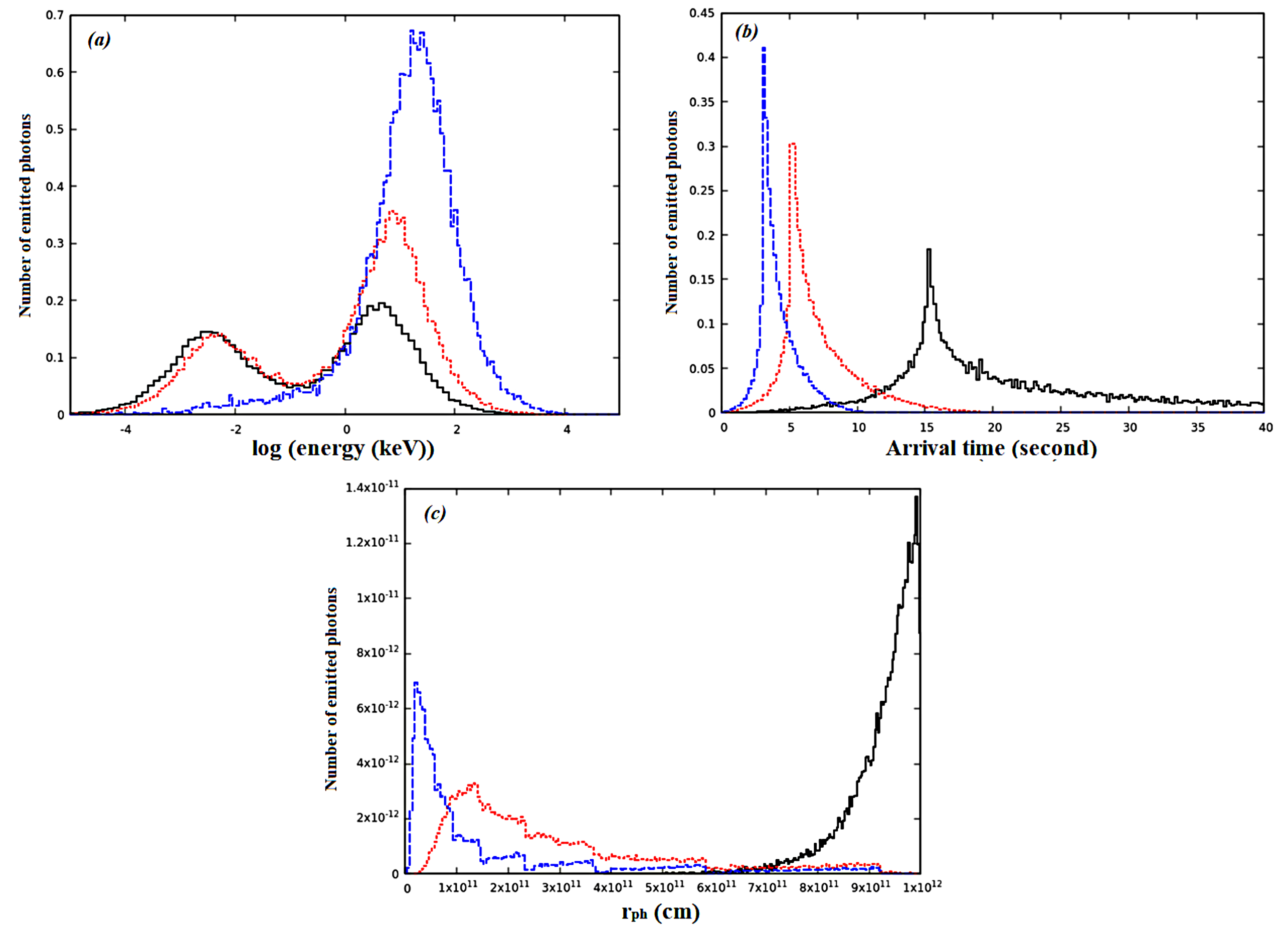}
\caption{Photon energy, arrival time and photospheric radius histograms (panels (a), (b) and (c), respectively) for three different values of the initial Lorentz factor, $\Gamma_{0} = 100$ (black solid spectra), $\Gamma_{0} = 300$ (red dotted spectra) and $\Gamma_{0} = 500$ (blue dashed spectra). In these plots, we took $\theta_{v} = 0$ and $\Gamma_{0} \theta_{j} = 1$.}
\label{fig:VariationGamma1}
\end{figure}

\begin{figure}
\centering
\includegraphics[width=1.\textwidth]{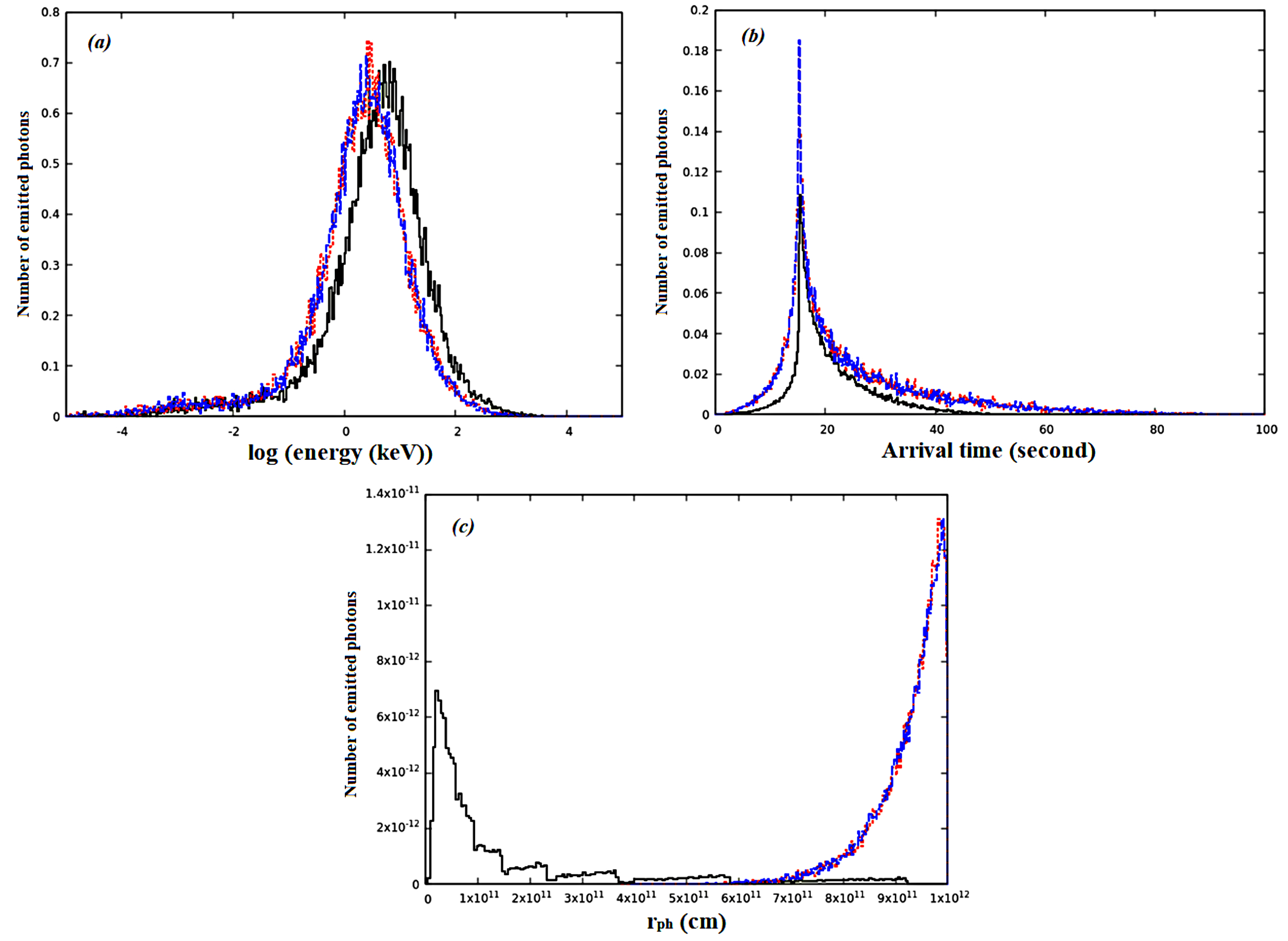}
\caption{Photon energy, arrival time and photospheric radius histograms (panels (a), (b) and (c), respectively) for three different values of the opening angle, i.e., $\theta_{j} = 1/\Gamma_{0}$ (black solid spectra), $\theta_{j} = 3/\Gamma_{0}$ (red dotted spectra) and $\theta_{j} = 10/\Gamma_{0}$ (blue dashed spectra). In these plots, we took $\theta_{v} = 0$ and $\Gamma_{0} = 500$.}
\label{fig:VariationThetaj1}
\end{figure}

The histogram of the arrival time calculated by Eq.\ref{ArrivalTime} is reported in the panel (a) of figure \ref{fig:TimeRph1}; it shows the distribution of times for a set of GRBs with a value of the initial Lorentz factor of $100$, in particular the time at which photons arrive at the observer regardless of the instant at which they were emitted. The most common time in this dataset is of around $22.9$ s, which is consistent with what one expects from GRBs theoretical models. In the panel (b) of this figure, we present the distribution of photospheric radii with the same value of the initial Lorentz factor. The most common photospheric radius in this histogram is of around $10^{12}$ cm, which is also consistent with the predictions of theoretical models.
To show the impact of the initial Lorentz factor, $\Gamma_{0}$, the opening angle, $\theta_{j} $, and the viewing angle, $ \theta_{v} $, on the photon energy, its arrival time and the photospheric radius histograms, we varied their values respectively in figures \ref{fig:VariationGamma1}, \ref{fig:VariationThetaj1} and \ref{fig:VariationThetav1}, that is $ \Gamma_{0} = 100$, $ 300$, $500$,  $\theta_{j} = few / \Gamma_{0}$ (with $few = 1$, $3$, $10$) and $\theta_{v} = few \times \theta_{j}$ (with $few = 0$, $1$, $2)$. 

By raising the photosphere's temperature, the Lorentz factor of a relativistic outflow significantly affects the photospheric emission, which has a number of implications on the observed photon spectrum. Due to the high motion rate of photons, their energy increases considerably along with the Lorentz factor. That's why the histogram in the panel (a) of figure \ref{fig:VariationGamma1} shows that as the initial Lorentz factor increases, i.e., $\Gamma_{0} = 100$, $300$, $500$, the corresponding peak energy of the photon increases, respectively as $4.46$, $7.22$, $22.38$ keV. As can be seen in this panel, for $\Gamma_{0} = 100$, $300$ the corresponding energy histograms exhibit low energy bumps respectively at $0.003$ and $0.005$ keV, while for $\Gamma_{0} = 500$, there is no clear evidence for a corresponding low energy bump.

Thus, higher values of the initial Lorentz factor mean that the photons have higher energies, which makes it more likely that the photons will interact with the matter contained in the outflow. This reduces the optical depth making it more difficult for the photons to escape from the outflow and reach the observer, leading to a smaller photospheric radius. This inverse relation between the Lorentz factor and the photospheric radius is confirmed in the photospheric emission theory by Eq.\ref{Rph1} and illustrated in our M C simulation by the panel (c) of figure \ref{fig:VariationGamma1} where the photospheric radius, $r_{ph}$, is equal to $9.933\times10^{11}$ cm, $1.334\times10^{11}$ cm and $1.9\times10^{10}$ cm respectively for $\Gamma_{0} =100$, $300$, $500$. Furthermore, for highly relativistic jets with large $ \Gamma_{0}$ values, $ r_{ph} $ may become smaller than some system length scale, leading to less pronounced low energy bump. 

The Lorentz factor plays a critical role in determining the presence and the strength of the fluctuations affecting the low energy range in GRB photospheric emission (see panel (a) of figure \ref{fig:VariationGamma1}). The panel (b) of this figure shows that the corresponding observed arrival times for $ \Gamma_{0} = 100$, $300$, $500$ are respectively $15.35$ s, $5.22$ s and $3.35$ s. This means that more energetic photons with higher Lorentz factor emitted from the front of the jet arrive earlier at the observer. Then, they have a reduced observed arrival time. The arrival time of photons emitted from the photosphere is influenced by the Lorentz factor in two ways: 
(i) It is delayed relative to the proper time of emission; as the Lorentz factor increases, time dilatation effects become more pronounced, which decreases the observed arrival time; the degree of the time dilatation depends on the Lorentz factor, with higher values leading to more significant time delay (see panel (b) of the figure \ref{fig:VariationGamma1}). 
(ii) Also, in the photospheric emission the photon's space coordinates and time in the rest frame of the photosphere are transformed into the observer's frame using the Lorentz transformation (see also the section\ref{code}) derived by the Lorentz factor, which then strongly affects the observed photon's arrival time.
The observed characteristics of the photospheric emission, such as the photon peak energy, the photospheric timescales and the position of the photosphere can be also affected by the jet's opening angle, $\theta_{j}$ (see figure \ref{fig:geometry} in section \ref{Introduction}). Larger opening angles imply that more photons are emitted farther relative to the jet axis. As a result, these photons experience less Doppler boosting, have longer paths to travel before reaching the observer along curved trajectories within the expanding jet material, and are featured by lower observed energies and longer observed arrival times. Furthermore, a broad jet with a large opening angle signifies that more material is involved in producing photon emission across larger solid angles around the jet axis. This increased material involvement leads to an expansion of the photosphere and an increase in its radius. These trends are consistent with the results of our M C simulation reported in figure \ref{fig:VariationThetaj1}. Indeed, with increasing the opening angle, i.e., $\theta_{j}= 1/500$, $3/500$, $10/500$ rad, the observed photon energy decreases respectively to $22.38$, $18.48$ and $16.59$ keV, while the low energy bump becomes (smoothly) more and more apparent (see the panel (a) of figure \ref{fig:VariationThetaj1}).
Besides, this increase in the angle $\theta_{j}$ leads, respectively, to increasing the photospheric radius to values of $1.9\times10^{10}$ cm, $9.8\times10^{11}$ cm, $9.9\times10^{11}$ cm (see the panel (c) of figure \ref{fig:VariationThetaj1}). In an other hand, one can observe in the arrival time histograms presented in the panel (b) of figure \ref{fig:VariationThetaj1} that an enhancement of the amount of photons occurs with increasing the opening angle, while the arrival time itself remains almost invariant. Thus, as can be seen in the panels (a), (b) and (c) of this figure, the effect of the opening angle on the photon's energy, its arrival time and on the photospheric radius is not particularly significant for $\theta_{j}$ values of $3/500$ and $10/500$. 
Now, observing the histograms shown in figure \ref{fig:VariationThetav1}, one notes that the viewing angle of a relativistic outflow significantly influences the observed properties of the photospheric emission. Relativistic jets have a Doppler boosting effect, making the photons more energetic (see the panel (a)) and arriving earlier (see panel (b)) at the observer for a smaller viewing angle. Also, a smaller apparent photospheric radius is observed from a closer viewing angle (see the panel (c)). 

\begin{figure}
\centering
\includegraphics[width=1.\textwidth]{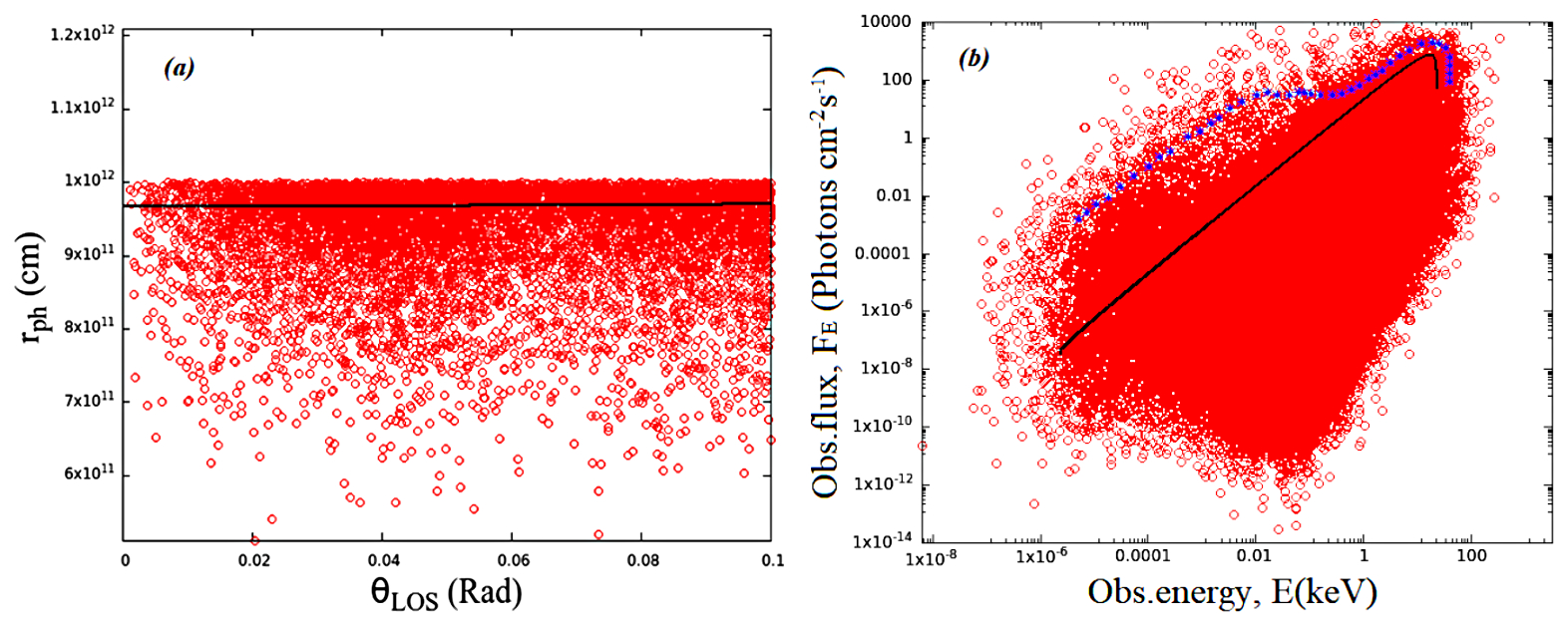}
\caption{Comparison of the results derived by M C simulation (red circles) and by numerical integration (black solid lines) for the photospheric radius (panel (a)) and for the observed photon flux (panel (b)) in the case of the basic plasma jet model. The M C simulation of the photon last scattering positions before reaching the observer was performed for Lorentz factor, $\Gamma_{0} = 100$, opening angle, $\theta_{j} = 10/\Gamma_{0}$, and viewing angle, $\theta_{v} = 0$. Colour degradation in panel (a) indicates M C simulated $ r_{ph} $ values with various probabilities, the most probable values being represented by the zone of maximum colour density. The curve in blue crosses in panel (b) marks the boundary of the general shape of the M C simulation results.}
\label{fig:Comparaison1}
\end{figure}

We have also written a numerical integration code for computing the photospheric radius and the observed photon flux using the Trapeze method \citep{Press}. In the case of the basic plasma jet model, these two quantities were calculated respectively by Eqs.\ref{Rph1} and \ref{Flux1}. In this code we have adopted the same free profile parameters as in the M C simulation and the analysis parts of the main code. We have set the total outflow luminosity to $L = 10^{52}$ $erg.s^{-1}$, the base outflow radius to $r_{0} = 10^{6}$ cm, the redshift to $z = 2$ and the luminosity distance to $d_{L} = 4.318$ Gpc. In the case of the basic plasma jet model, the calculations were performed for $\Gamma_{0} = 100$, $\theta_{j}=10/\Gamma_{0}$ and $\theta_{v} = 0$. The resulting spectra are reported in figure \ref{fig:Comparaison1} together with those derived by M C simulation. Thus, the results derived by these two different methods for the photospheric radius (shown in panel (a)) and the observed photon flux (reported in panel (b)) appear to be mutually very consistent, those derived by numerical integration being in best agreement with the most probable values obtained by M C simulation. One can also observe in this figure that greater ranges of values for the two computed observables are seemingly derived by M C simulation in comparison to the ranges of values inferred by numerical integration. This is due to the difference in nature between these two computational methods : while the latter method simply refers
to a direct mathematical calculation, the former one, based on random numbers and probability distribution models, produces a larger number of estimated possible values. 

As a function of the line of sight angle $ \theta_{LOS} $, the M C simulated values of the photospheric radius exhibit low dispersion and stay clustered around the numerically integrated line. This indicates that, for small LOS angles, the M C simulation accurately depicts $r_{ph}$, which remains nearly constant with values of around $10^{12}$ cm for $ 0 < \theta_{LOS} < 0.1 $ rad.
Besides, concerning the observed photon flux, the associated spectrum generated by numerical integration indicates a low-energy peak at around $19$ keV, a value which is also consistent with our result derived by M C simulation. In addition, a bump is pointed out at very low photon energies by M C simulation that is not similarly reproduced by numerical integration.

The observed energy spectra in the GRB prompt emission are usually well described by a Band function. Then, we have fitted such a function to the numerically integrated spectrum shown in panel (b) of figure \ref{fig:Comparaison1} (black solid line) using the Astropy (https://www.astropy.org) and Scipy (https://www.geeksforgeeks.org/scipy-curve-fitting/).optimize.curve fits packages, Python tools libraries that provide functionalities for various astrophysical tasks including time series analysis and fitting of predefined or customized functions to photon spectrum. We obtained best-fit values of $\alpha \approx 0.45$ and $\beta \approx -2.6$ respectively for the low and high energy photon indices. The latter value of the high-energy photon index agrees very well with the value of $-2.5$ reported in \cite{Ito2014} and \cite{Ito2024}, while our result for $\alpha $ is very consistent with the common value of $ 0.5 $ reported by \cite{Lundman}, \cite{Meng2018} and \cite{Deng} in the case of the isotropic plasma jet. These photon index values, theoretically expected for the spectral shape of a photospheric emission component in GRBs, indicate a rising power-law at low energies and an extremely steep cut-off at high energies. But such a function clearly does not match a blackbody spectrum which is a narrow continuum characterized, instead, by a low energy photon index  $\alpha = 1$. In addition, the blackbody spectrum was found to be in sharp contrast with the M C simulated one (red dashes in panel (b) of figure \ref{fig:Comparaison1}) that appears to be significantly broader. Consequently, one cannot explain the thermal photon emission by a simple blackbody model. The energy loss close to the photosphere has been suggested \citep{Beloborodov13, Beloborodov, Vurm} as an alternative way that alters the local comoving photon spectrum for making the photospheric model consistent with observations. In this case, multiple Compton scattering near the photosphere can significantly broaden the observed spectrum and alter its low-energy part, causing spectral distortions and anisotropies in comparison to a pure blackbody distribution.\citep{aksenov}.

\subsection{Structured plasma jet model}\label{structured results} 
\begin{figure} 
\centering
\includegraphics[width=1.\textwidth]{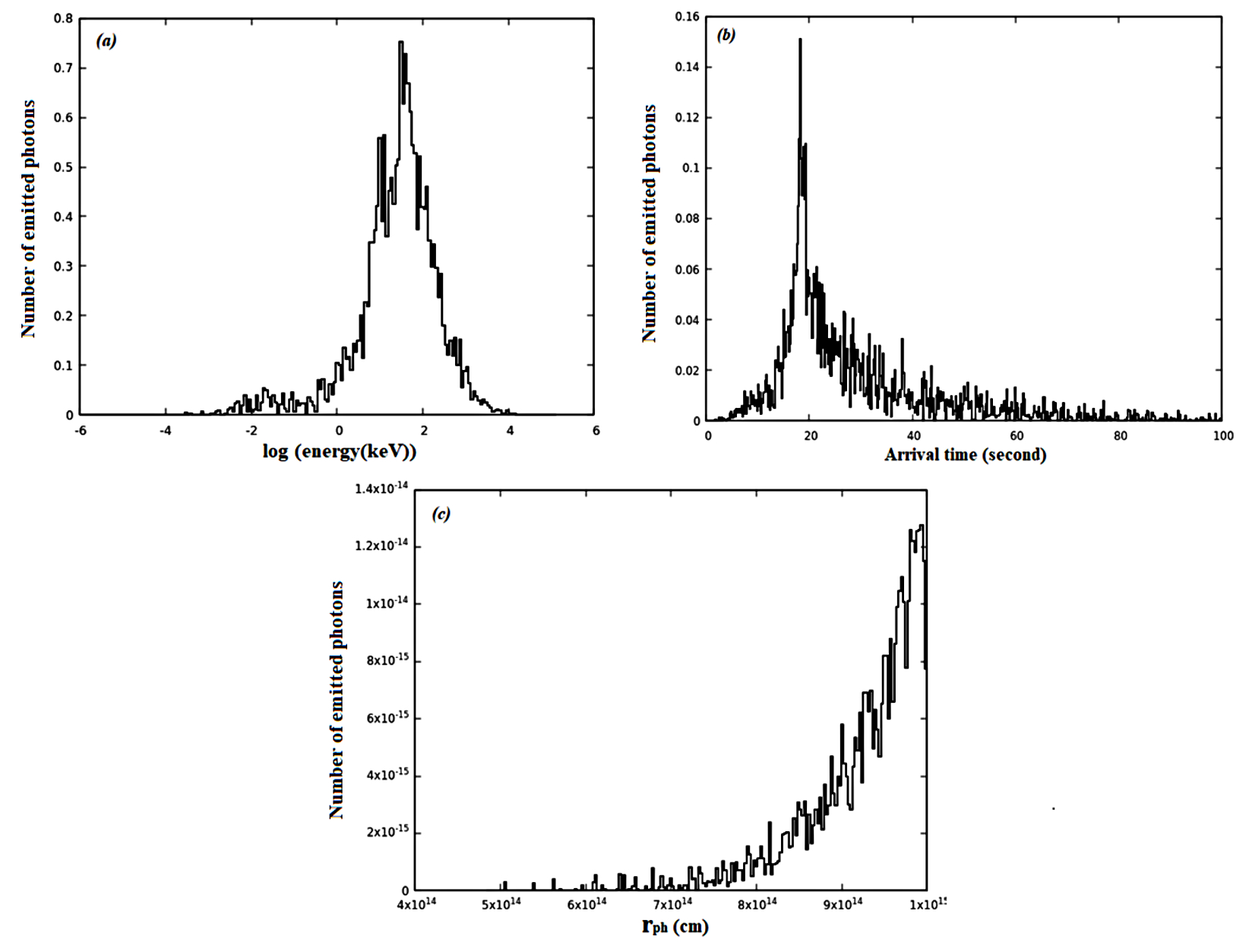}
\caption{Histograms of the photon energy, its arrival time and the photospheric radius (panels (a), (b) and (c), respectively) for initial Lorentz factor, $\Gamma_{0} = 100$, opening angle, $\theta_{j} = 1/100$ rad, and base outflow radius, $ r_{0} = 10^{8}$ cm. A viewing angle, $\theta_{v} = 0$, is assumed. A power-law index, $p = 1$, was taken in the calculation of the angular dependence of the Lorentz factor via Eq.\ref{gammatheta}.}
\label{fig:EnergyTimeRph2}
\end{figure}
In the case of the structured plasma jet model the photon energy, its arrival time and the photospheric radius histograms are presented in figure \ref{fig:EnergyTimeRph2}. The calculation was performed by setting the base outflow radius to $r_{0} =10^{8}$ cm, the power-law index to $p = 1$, $\Gamma_{0} = 100$, $\theta_{j} =1/100$ rad and $\theta_{v} =0$. The observable properties of the photospheric emission are strongly influenced by the angular distribution of the Lorentz factor. 

Two photon peak energies can be distinguished in the photon energy histogram (see panel (a) of this figure) at around $32.35$ and $10$ keV, as well as a small low energy bump at $0.02$ keV. The first peak energy value is validated through the observation by the Fermi-GBM of some GRBs tabulated in \cite{Nava}, registering a close peak energy value for these GRBS : the GRB $100307A$, GRB $090815B$ and GRB $090824A$ showing peaks at energy values of $32.53 \pm 4.16$ keV, $32.36 \pm 1.43$ keV and $30.33 \pm 1.3$ keV, respectively, and the GRB $111228A$ exhibiting multiple peaks with the main peak growing at $34 \pm 0.1$ keV \citep{VonKielen}. The first and second peaks in panal (a) are most likely generated by photons released, respectively, from the jet's inner and outer areas delimited by angles of $(\theta < \theta_{j})$ and $(\theta_{j} <\theta < \theta_{e})$, respectively, while the small low energy bump can be explained, as in the preceding subsection, by multiple Compton scattering effects on the photon energy spectrum.

The arrival time of photons emitted from the outer region and close to the envelope is geometrically delayed, up to $18$ seconds (see panel (b) of figure \ref{fig:EnergyTimeRph2}), compared to the case of the basic plasma model. As for the photospheric radius (see panel (c)), its value is increased relative to the basic plasma jet model, up to $10^{15}$ cm and can even reach a larger value of $10^{16}$ cm.

\begin{figure}
\centering
\includegraphics[width=1.\textwidth]{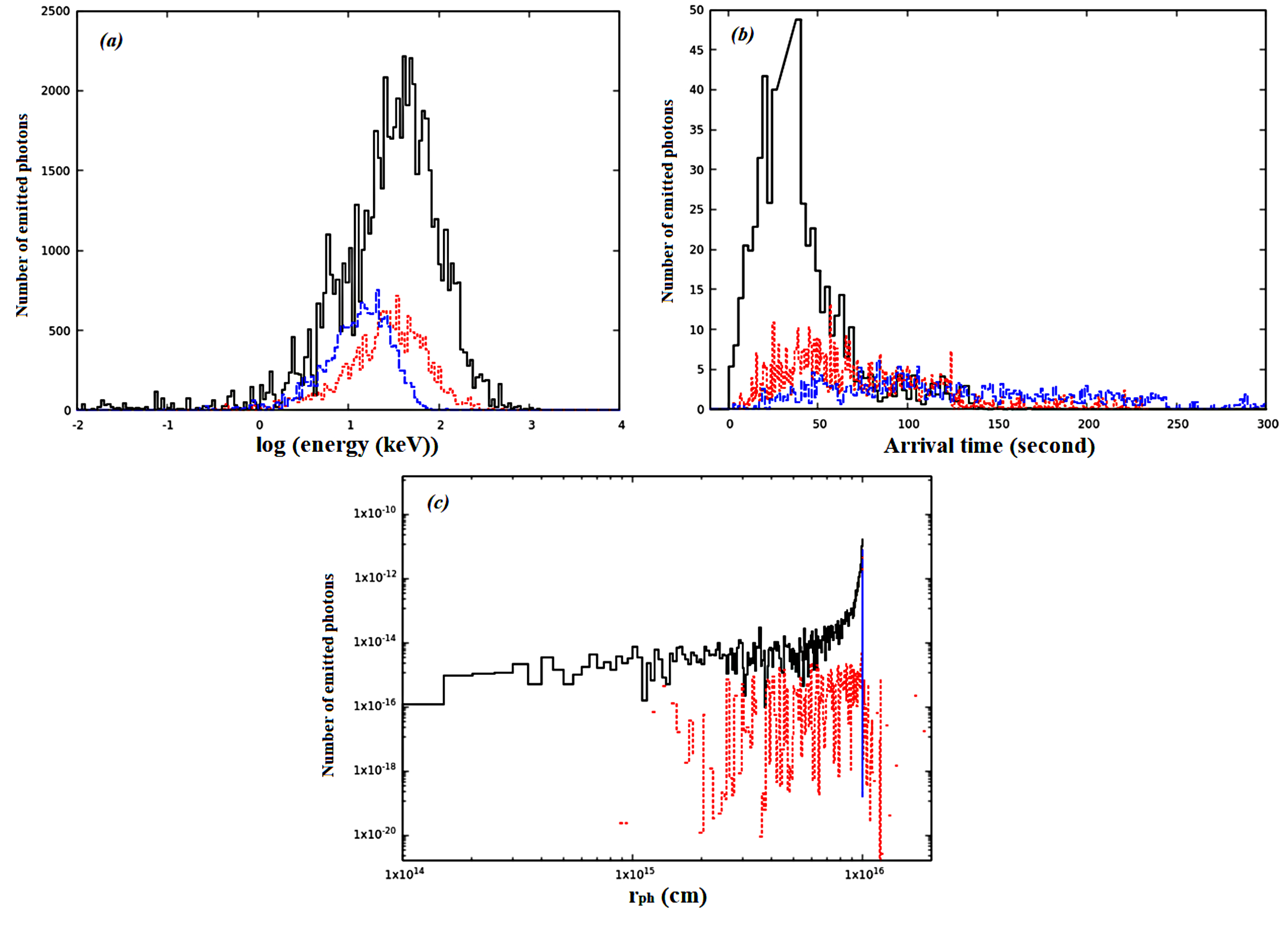}
\caption{Histograms of the photon energy, arrival time and photospheric radius (panels (a), (b) and (c) respectively) for three different values of the power-law index $p = 1$, $2$, $4$  (black solid, red dotted and blue dashed photon spectra, respectively). These plots were generated by taking $\Gamma_{0} = 500$, $\theta_{j} = 1/\Gamma_{0}$ and assuming a zero viewing angle, $\theta_{v} $. }
\label{fig:VariationIndexP}
\end{figure}

Investigating the behaviour of the photospheric emission properties under the variations of the Lorentz factor, the opening angle and the viewing angle, we found that they undergo similar effects as in the case of the basic plasma model discussed in the preceding subsection (see Figures \ref{fig:VariationGamma1}, \ref{fig:VariationThetaj1}, \ref{fig:VariationThetav1}). 
Furthermore, as discussed in section \ref{jet models}, the study of the radiative transfer in a structured relativistic plasma jet, an angle-dependence description of its properties is needed with considering the angular profiles of the associated free parameters, such as the Lorentz factor described by a power-law function of the angle of index p (see equation \ref{gammatheta} and figure \ref{fig:gammaProfile}) over the plasma jet's outer region. Thus, figure \ref{fig:VariationIndexP} reports histograms of the photon energy, its arrival time and the photospheric radius with varying the power-law index, i.e., $p = 1$, $2$, $4$. Here, we set the initial Lorentz factor to $\Gamma_{0} = 500$, the opening angle to $\theta_{j}= 1 /\Gamma_{0}$ and the viewing angle to $\theta_{v}= 0$.

Since $85\%$ of the emitted photons originate from the outer jet area and the envelope, the power-law index $p$ significantly affects the emission characteristics. As described earlier in section \ref{jet models}, the angular distribution of the Lorentz factor implies a reverse relationship between the Lorentz factor and the index $p$. This leads to an opposite effect of the power-law index $p$ on the emission characteristics relative to the Lorentz factor. Consequently, as can be observed in panel (a) of figure \ref{fig:VariationIndexP}, the photon peak energy decreases with increasing the index $p$, and we thus obtained peak energy values of $50$ keV, $31.6$ keV, and $15.8$ keV respectively for index $p = 1$, $2$ and $4$.
In contrast, since the photon arrival time is proportional to the power-law index $p$, we obtained values of $38$, $47$ and $52$ seconds, respectively, for these values of index p (see panel (b) of figure \ref{fig:VariationIndexP}). Nevertheless, the photospheric radius remains practically constant with varying the power-law index p, coming out to a value of $10^{16}$ cm, as can be seen in panel (c) of figure \ref{fig:VariationIndexP}.

\begin{figure}
\centering
\includegraphics[width=1.\textwidth]{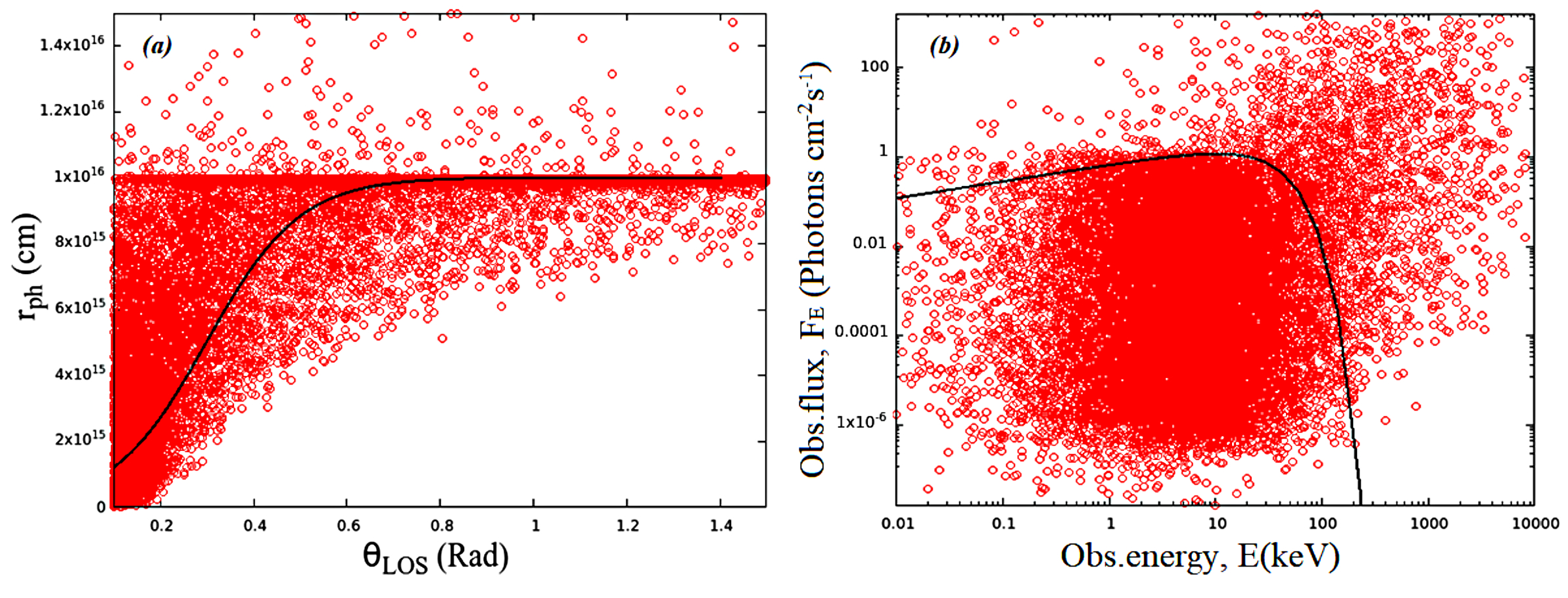} 
\caption{Comparison of the results derived by M C simulation (red circles) and by numerical integration (black solid lines) for the photospheric radius (panel (a)) and the observed flux (panel (b)) in the case of the structured plasma jet model. The values derived by numerical integration clearly appear to be in good agreement with the most probable values obtained via M C simulation. Both trwo computations were performed for Lorentz factor, $\Gamma_{0} = 100$, opening angle, $\theta_{j} = 10/\Gamma_{0}$, viewing angle, $\theta_{v} = 0$ and a power-law index, $p = 2$.}
\label{fig:Comparaison2}
\end{figure}

We determined the photospheric radius by Eq.\ref{Rph2} and the measured photon flux by Eq.\ref{flux2} using the same numerical integration code and the procedure as in the case of the basic plasma model. The derived results for the two observables with setting the free structure parameters to  $\Gamma_{0} = 100$, $\theta_{j} =10/\Gamma_{0}$, and $\theta_{v} = 0$ and $p = 2$ are shown in figure \ref{fig:Comparaison2} together with those obtained by M C simulation for the same values of the free parameters (those for the photospheric radius being displayed in panel (a) of this figure, while those for the photon flux are reported in panel (b)). 

Similarly as in the case of the basic plasma jet model (see section \ref{basic results}), the results derived following the two different methods are thus found to be mutually quite consistent, those derived by numerical integration being in best agreement with the most probable values generated by M C simulation. Notice again that due to the difference in nature between these two computational methods, this figure also shows greater ranges of values derived by M C simulation for the two observables than the ranges of values inferred by numerical integration. 

The simulation results show a noticeable rise in the photospheric radius with the line of sight angle, $ \theta_{LOS} $. At small angles ($0<\theta_{LOS} < 0.8$ rad), the photospheric radius exhibits a significant variability, spanning several orders of magnitude. At larger angles ($\theta_{LOS} > 0.8$ rad), however, the $ r_{ph}$ values converge to $ 1.05\times10^{16}$ cm. Therefore, the photosphere clearly appears to have a strong angular dependance with smaller radii observed near the line of sight and larger radii at wider angles. This observed aspect reveals that the photosphere is anisotropic, non-spherical and potentially quasi-thermal, which reflects the complex and dynamic nature of the jet's outer regions and implies that the shape of the photosphere is strongly modified under the geometrical effects \citep{beloborodov2017, song2022}.

Concerning the photon flux results reported in panel (b) of the figure \ref{fig:Comparaison2}, one observes that the resulting M C simulated spectrum, showing to be broad and exhibiting a peak energy value at around $32$ keV, does not obviously resemble to a narrow blackbody function.

As in the case of the basic plasma jet model, we fitted a Band-law function to the numerically integrated spectrum using the Astropy and Scipy.optimize.curve fit libraries. We obtained a low-energy photon index value of $\alpha\approx -0.55$ consistent with the value of $\alpha \approx -0.625$ reported by \cite{Lundman} for the same value of $p = 2$, and a high-energy photon index value of $ \beta \approx -3 $ in excellent agreement with the value ($ \beta \approx -3 $) found by \cite{Meng2024}. For index $p = 0$, the outflow profile is spherically symmetric with a Lorentz factor, $\Gamma$, lower than the initial Lorentz factor, $\Gamma_{0}$, for $(\theta_{j} < \theta < \theta_{e})$. An excellent agreement was obtained between the M C simulated and the numerically integrated spectra for index $p =2$, as can be observed in panel (b) of figure \ref{fig:Comparaison2}. Both two spectra thus exhibit a large broadening, which is evidence of important contributions of photons emitted from the outer jet region ($ > 80\%$) and, consequently, of the influence of the geometrical effects on the GRB photospheric emission, due to the angular dependence of the Lorentz factor. Also, the expected bump at very low photon energies in the M C simulated spectrum is masked by the spectral broadening in the panel (b) of figure \ref{fig:Comparaison2}.

Comparing the photospheric radius values and the photon spectra derived from the two postulated distinct plasma jet models, it appears evident that the geometrical effects significantly broaden the GRB spectral shapes. In the case of the basic plasma jet model, the derived low-energy photon index $ \alpha $ amounts to $0.45$, whereas in the structured plasma jet model, $ \alpha$ shifts to $-0.55$. In addition, the photospheric radius is also notably affected, increasing from $ 10^{12}$ cm in the former model to $ 10^{16}$ cm in the latter one. Therefore, deeper and more complex interpretation of the GRB photospheric emission is obtained by including geometrical factors in structured plasma jet models. With the photosphere likely being no longer a perfect sphere as in the case of the basic plasma jet and varying in density and other characteristics in different directions within structured plasma jets, the latter models should provide a more accurate representation of the relativistic system. E.g., the jet's energy and dynamics can then be calculated with greater accuracy enabling one to get important knowledge about the outflow collimation and its structure, both two properties being of crucial significance for better mastering the physics of GRBs. Thus, more sophisticated simulations could be performed for gaining deeper insights on these powerful cosmic events, which can be attained only through structured plasma jet models that take geometrical effects into account.

Therefore, by varying the free outflow parameters in our M C simulation code based on the photospheric emission model and in our numerical integration computations, we derived various results for the properties of GRBs in very good agreement with the predictions of this model, with values from previous works and with GRB observational data. This then widely validates the robustness of our M C code and its performances.  
 
\section{Data interpretation} \label{interpretation}
In order to validate the computational techniques used in our M C simulation and analysis code of GRBs (see Appendix), we have compared results derived by the latter, in particular the GRB spectra generated over the photon energy range of $E_{\gamma}$ $< 100$ keV, to counterparts observed by the GBM onboard the FERMI space telescope. For this purpose, we have explored the GRB observational data registered in the Fourth Fermi-GBM GRB Catalog \citep{VonKielen} that presents observed bursts over the time period of $10$ years, from July $12$, $2008$ to July $11$, $2018$. We counted $21$ interesting GRBs based on their low energy range, the majority of them having peak energies in the range extending from $10$ to $70$ keV, while some exhibiting a multiple peak structure in the energy band of $E_{\gamma}$ $< 100$ keV. 

\begin{figure}
\centering
\includegraphics[width=1.\textwidth]{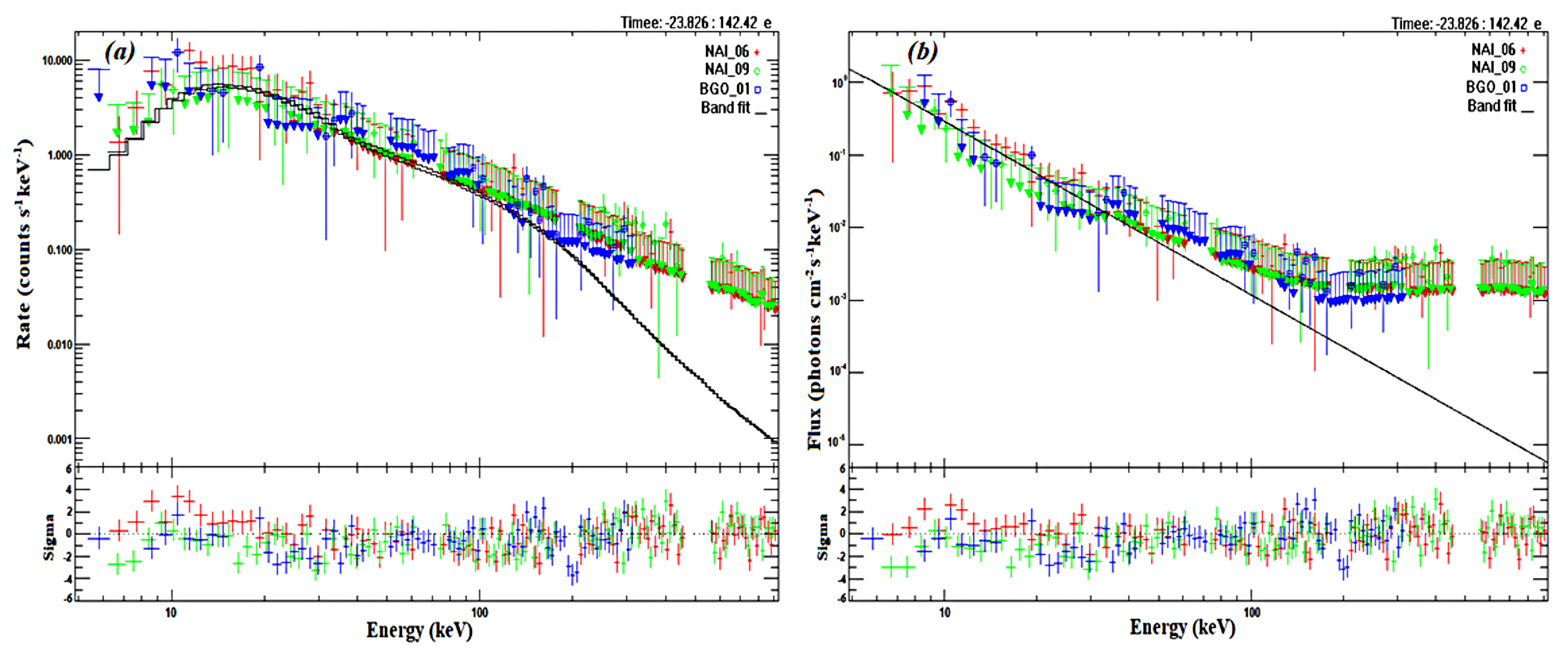}
\caption{Observed and fitted energy spectra for the GRB $090706$ expressed in terms of photon rate (panel a) and photon flux (panel b); the black solid line represents the spectrum generated in our Band analysis via the RMFIT software, while the observed spectra by the $NAI-06$, $ NAI-09 $ and $ BGO-01 $ detectors are shown, respectively, by green circles, red squares and blue crosses.}
\label{fig:GRB090706}
\end{figure}

\begin{figure}
\centering
\includegraphics[width=1.\textwidth]{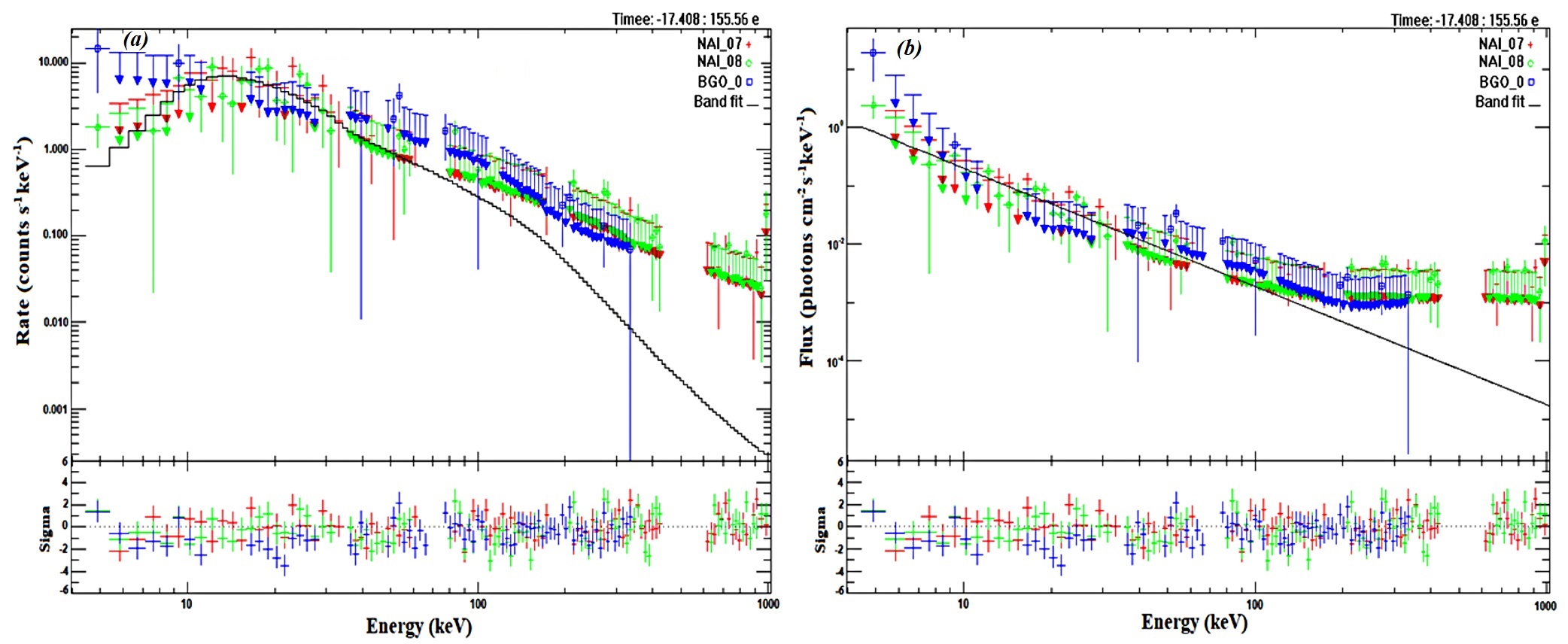} 
\caption{Same as in figure \ref{fig:GRB090706} but for the GRB $GRB090807$ with the green circles, the red squares and the blue crosses now representing, respectively, the observed spectra collected by the $NAI-07$, $ NAI-08 $ and $ BGO-01 $ detectors.}
\label{fig:GRB090807}
\end{figure}

We selected two GRBs for comparisons to the outputs from our M C simulation code and numerical integration analysis : the GRB $090706$ and the GRB $090807B$ that we believed to be among the most suitable ones. This selection was made based on the correlations between the values of the photon peak energies and the $ \alpha $ and $ \beta $ low and high energy photon indices for these bursts and the values of these parameters we derived by our M C simulation and numerical integration. We have extracted bursts informations from the GRB Coordinates Network (GCN)\footnote{https://gcn.gsfc.nasa.} and entered them into the browsable catalogs accessible online at HEASARC and FERMIGBRST\footnote{Fermi-GBM burst catalog at HEASARC (https://heasarc.gsfc.nasa.gov/w3browse/fermi/fermigbrst.html)} in order to download the burst files of interest. We performed a single Band analysis of the observational spectra for these two GRBs within the RMFIT software $(version 3.3pr7)$\footnote{https://fermi.gsfc.nasa.gov/ssc/data/p7rep/analysis/rmfit/}. For this purpose, we selected the most illuminated NaI and BGO detectors having an angle lower than $80°$ between the photons emitting source and the normal to the detector, i.e., the $NAI-06$, $NAI-09$ and $BGO-01$ detectors for the GRB $090706$ and the $NAI-07$, $NAI-08$ and $BGO-01$ ones for the GRB $090807B$. We used the CSPEC and TTE files with respective time resolutions of $1.024$ s and $0.064$ s. The obtained results are presented in the figures \ref{fig:GRB090706} and \ref{fig:GRB090807} where the observed spectra and corresponding fitted ones are expressed in photon rate units (i.e., in $counts.s^{-1}.keV^{-1}$ , panels a) and in photon flux units (i.e., in $photons.cm^{-2}.s^{-1}.keV^{-1}$ , panels b).

The GRB $090706$ is a soft burst in the Galactic plane notably exhibiting a distinctive profile with one discernable pulse in the photon low-energy range; its peak energy was registered at $17.4 \pm 0.64$ keV. Our Band spectral analysis of its data with the RMFIT program yielded a low-energy index of $\alpha = 0.42 \pm 0.5$ and a high-energy index of $\beta = -2.66 \pm 0.09$ (see the figure \ref{fig:GRB090706}). These spectral properties match very well the results derived by our M C simulation and numerical integration in the case of the basic plasma jet model, i.e., one peak energy at around $ 19 $ keV, a low-energy photon index of $ \alpha= 0.45 $ and a high-energy photon index of $ \beta = -2.6 $ for values of the outflow free parameters of $ \Gamma_{0}=100 $, $\theta_{j}=10/100 $ rad and $ \theta_{v}=0 $ (see subsection \ref{basic results}). 
 
The GRB $090807$ spectrum consists of only one pulse with a peak energy at $36.8 \pm 5.1$ keV; Our Band spectral analysis with the RMFIT software yielded values of $\alpha = -0.61\pm 0.3$ and $ \beta = -2.89 $ respectively for the low-energy and high-energy photon indices (see the figure \ref{fig:GRB090807}). These values are also in fairly good agreement with the results derived by our M C simulation and numerical integration in the case of the structured plasma jet model, i.e., one peak energy at around $32$ keV and low and high-energy photon indices of $\alpha = -0.5$ and $ \beta = -3 $, respectively for values of the outflow free parameters of $\Gamma_{0} = 100$, $\theta_{j}=10/100 $ rad, $ \theta_{v}=0 $, and index $p =2$. 

Besides, in the case of the structured plasma jet model and for index $p = 1$, we derived results by M C simulation and numerical integration consistent with the spectral properties of the GRB $090708A$ and the GRB $080806$ observed by the Fermi-GBM. These two GRBs are both registered with the same peak energy of about $50$ keV and respective values of $\alpha$ = $-0.80 \pm 0.31 $ and $-0.75 \pm 0.10$ of their associated low-energy photon index, and with respective values of $\beta = -2.58\pm 0.32$ and $ -2.27\pm0.05 $ of their high-energy photons index \citep{Nava}.

Therefore, one can conclude that both two plasma jet models considered in this study provide a good interpretation of the observed spectra for the selected four GRBs (and likely several other GRBs), which reveals the force of our M C simulation code based on the photospheric emission model to well account for the GRB observational data when one applies appropriate outflow structure parameters. 
Besides, both two plasma jet models considered in this work are featured by photospheric radii lying within the range extending from about $10^{12}$ to $10^{16}$ cm (see the panels (a) in Figures \ref{fig:Comparaison1} and \ref{fig:Comparaison2}) and can even reach $10^{18}$ cm for index $p=1$, in fair consistency with previous values reported in the literature \citep{Li21, aksenov, Lundman, Peer}.

Finally, an alternative approach to the Band function and the cut-off power-law spectral analyses can be used for interpreting the GRB emission and observational data. This approach, used notably in recent studies by \cite{Meng2018, Meng2019, Meng2024}, consists in directly fitting theoretical models based on the photospheric emission to GRB observed spectra or light curves and leads to results in good agreement with those derived via these analyses . 

\section{Summary, conclusion and perspectives}\label{conclusion} 
We have carried out a comprehensive study of the optically thick GRB photospheric emission model predicting a blackbody or quasi-blackbody spectrum and suggested earlier for explaining a still non elucidated thermal component of the observed GRB prompt emission spectrum.
We have first outlined the corresponding theory for enlightening the subsequent description of our M C simulation code and its application. We considered a basic, isotropic relativistic plasma jet where the bulk Lorentz factor is taken constant and a structured plasma jet supposing this factor angle-dependent and obeying a decreasing power-law of the angle of index $p$ over the outflow outer region, which leads to complex angular dependencies of the photospheric emission properties. 
The provided M C code, written in C++ language, is mainly composed of a simulation part for accurately tracking the photon propagation via a multiple Compton scattering mechanism below the photosphere and an analysis part for exploring the photospheric emission properties (the energy, arrival time and observed flux (spectrum) of the simulated escaped seed photons, and the photospheric radius). Besides, in order to independently compute the latter two observables by numerical integration, a specific software has been also written and incorporated into the main M C code, thus constituting a third part of it. Notice that we are mainly concerned by the photon low-energy region of $E_{\gamma}$ $ < 100$ keV far below the characteristic peak energy around $1$ MeV of the GRB prompt emission spectra.

The derived results are reported, confronted to GRB observations made by the Fermi-GBM mission and discussed in details. They show large consistencies with the predictions of the photosphetric emission model, with GRB observational data and with previous counterparts from the literature. Thoroughly investigated variations with the outflow free parameters (Lorentz factor, opening angle, viewing angle and power-law index $p$) of the photospheric emission properties show that the latter are especially sensitive to the Lorentz factor that plays a crucial role in determining the presence and strength of the geometrical effects manifesting themselves through substantial broadenings of the emission spectra. The latter are found to be remarkably structured and featured by multiple peak energies with the generation of bumps at very low photon energies. In particular, the derived values by M C simulation and numerical integration for the photospheric radius and the observed photon spectrum show very good mutual agreements in the cases of both two considered plasma jet models. 

Referring to the figure \ref{fig:Comparaison1} associated with the basic plasma jet model, the numerically integrated and MC simulated values of the photospheric radius are found to be in very good agreement, amouning to $ 10^{12}$ cm for $ 0 < \theta_{LOS} < 0.1 $ rad. Similarly, referring to figure \ref{fig:Comparaison2} for the structured plasma jet model, the $ r_{ph} $ values derived by the preceding two methods, amounting to $ 10^{16}$ cm for $ \theta_{LOS} > 0.8 $ rad, are also found to be very well matching. Besides, the values of $ r_{ph} $ inferred by the two methods exhibit a similar variability for $0.1 < \theta_{LOS} < 0.8 $ rad.
These calculations were performed for initial Lorentz factor, $\Gamma_{0} = 100$, opening angle, $\theta_{j}=10/\Gamma_{0}$, viewing angle, $\theta_{v} = 0$, for both two plasma jet models and, in addition, a power-law index $p = 2$ in the case of the structured plasma jet model. Thus, the overall values of $ r_{ph} $ for both two plasma jet models lie within the extended range of $10^{12} < r_{ph} < 10^{16}$ cm. Notice that the upper value of $10^{16}$ cm derived by M C simulation for $ r_{ph} $ with index $p = 2$ in the case of the structured plasma jet model could be even much larger for larger values of the index p. Besides, referring to the same two preceding figures, the values of the observed photon flux vary within the interval extending from $ 1$ up to $ 655.25$ $ erg$ $ cm^{-2} s^{-1} erg^{-1}$ for both two plasma jet models.

In the case of the basic plasma jet model and for the same preceding values of the free parameter cited above, the M C-simulated spectrum shows a prominent peak at a photon energy of about $19$ keV found to be very consistent with GRB observations at energies $ < 100 $ keV, as well as a bump growing at very low photon energy of $E_{\gamma}$ $\sim 0.013$ keV, and significant broadening in both cases (see figures \ref{fig:energy1}, \ref{fig:VariationGamma1}, \ref{fig:VariationThetav1} and \ref{fig:Comparaison1}). The very low-energy bump is also evidenced in figure \ref{fig:energyTime} reporting the variation of the photon energy versus the photon arrival time. As discussed in section \ref{basic results}, these observations reflect a GRB photospheric emission at very high temperature amounting to about $10^{9} K$, featuring strong bursts generated by the collapse of massive stars. Notice, however, that the very low-energy bump was not similarly reproduced by numerical integration, likely because of our negligence of the contribution to the observed photon spectrum due to the impact of the surrounding medium. Consequently, considering this contribution in the computation should expectedly provide a more accurate representation of the observed photon spectrum.

Similar results have been derived in the case of the structured plasma jet model. The observed photon spectrum also shows substantial broadening and a structured shape with multiple photon peak energies, i.e., two pronounced peaks at around $32$ and $10$ keV likely corresponding to photon emissions respectively from the jet's inner and outer regions, and a small bump at very low photon energy of $0.02$ keV (see figures \ref{fig:EnergyTimeRph2}, \ref{fig:VariationIndexP} and \ref{fig:Comparaison2}). The energy of the first, most prominent peak was found to match very well the reported values for several GRBs observed by the Fermi-GBM mission. 

Notice, however, that significant differences also distinguish the photospheric emission properties in the basic plasma jet and the structured plasma jet. In the latter model, higher values of the photospheric radius and the observed photon peak energies are obtained, the arrival time of photons emitted from the outer region and close to the envelope is significantly more delayed (up to $18$ seconds, see figure \ref{fig:EnergyTimeRph2}), while the photospheric emission properties are strongly impacted by the variations of the angle-dependent Lorentz factor and the associated power-law index $p$ (see figure \ref{fig:VariationIndexP}). The pointed out considerable broadening of the observed spectral shapes stem from two sources : (i) the multiple Compton scattering mechanism below the photosphere affecting the results from both two plasma jet models \citep{Beloborodov13, Beloborodov, Vurm}, \citep{aksenov} and (ii) the geometrical effects \citep{Lundman} that severely broaden the observed photon spectrum in the case of the structured plasma jet model due to the important radiation from the jet's outer region (i.e., for $\theta_{j} < \theta < \theta_{e}$) featured by a decreasing bulk Lorentz factor, $\Gamma\sim\theta^{-p}$ (see \ref{structured plasma model}).

In an other hand, fitting Band functions to the numerically integrated spectra, we obtained best-fit values for the low-energy photon index of $\alpha \approx 0.45$ and $\approx -0.55$ respectively in the cases of the basic and the structured plasma jets. These values are very consistent with those featuring most typical observed GRBs, with values from the previous works and with theoretical predictions of the GRB photospheric emission model. Similarly, best-fit values of $\beta \approx -2.6$ and of $ \approx -3 $ for the high-energy photon index have been derived respectively in the cases of the two considered plasma jet models, which are also very very consistent with previous counterparts from the literature and theoretical predictions. However, these results pointing to soft and broad spectra reflect associated power-law functions rising at low energies and exhibiting steep cut-off at higher energies, which is incompatible with the hard, narrow shape featuring a blackbody spectrum obeying a power-law function with low-energy photon index $\alpha = 1$. 

In conclusion, by varying the outflow structure parameters in the M C simulation and data analysis computations with our M C code, we have derived various results for the main properties of the GRB photospheric emission, also confirmed by numerical integration in the cases of the photospheric radii and the observed photon spectra. These results were found in fairly good agreement with numerous GRB observational datasets, with counterparts from previous works and with the photospheric model predictions, which led us to perform detailed further Band analyses within the RMFIT software of selected GRB observed spectra (see section \ref{interpretation}). The outcomes point out the strength of the photospheric emission model to interpret the observations and well account for the GRB spectral properties far below the main peak energy around $1$ MeV in the GRB prompt emission phase without having recourse to invoke multiple radiative processes. Therefore this model, featured by a low energy photon index $\alpha \approx 0.5$ associated with most observed GRBs, stands to be a more popular than expected and efficient scenario for explaining the presence of a thermal component in the prompt emission spectra in both cases where one assumes a basic relativistic plasma jet or a more realistic and complex structured jet featured by strongly angle-dependent photospheric emission properties.
Thus, several key findings highlight the results obtained in the current study that confirm, complete and/or extend previous results collected by several groups of workers, notably by \cite{Lundman} (see also the Sections \ref{results} and \ref{interpretation}). Especially, we emphasize the importance of the following ones : 
(i) The remarkable variability of the photospheric radius versus the LOS angle in the case of the structured plasma jet model where the photosphere becomes strongly angle-dependent, anisotropic, non-spherical and potentially quasi-thermal under geometrical effects.
(ii) The great impacts of the multiple Compton scattering and geometrical effects on the simulated photon spectra and related spectral fluctuations over the very low-energy regime of the GRB prompt emission phase in the cases of the basic and structured plasma jet models.
(iii) The predicting strength of our M C simulation code for reproducing the observational data for a large number of observed GRBs, which clearly validates this code. 

Nevertheless, the observed GRB prompt emission spectrum involving a thermal component could be further constrained by considering the interplay of multiple radiation mechanisms including photospheric emission, sub-photospheric energy dissipation, comptonization of the thermal spectrum and the magnetic reconnection, the first three ones likely contributing to the observed considerable spectral broadening. Then, complementary research is needed for accurately elucidating the role of these radiation processes far at sub-MeV energies. In addition, to fully reproduce the GRB observational data, other more powerful structured plasma jet models could be considered in our M C simulation code. The latter has at least two striking advantages : (i) its analysis part, conceived as entirely separated from the modeling of the involved radiation mechanism, will remain the same as new plasma jet models are postulated in its simulation part and (ii) the polarization parameters are pre-implemented in this code, which would enable one to simultaneously consider non-polarized and/or polarized relativistic plasma jet models. the relative weights of several radiation mechanisms could be even quantified with identifying which model predominates in reproducing the GRB observational data. 
In another hand, the present M C code could be further improved to include the evaluation of various correlations studied previously via hydrodynamical and/or M C simulations of the GRB photospheric emission and still debated in the literature \citep{Lazzati13, parsotan2018, parsotanLazzati2018, Ito2019, Ito2021, Ito2024}. In particular, one could explore, e.g., the correlations between the peak energy, the peak luminosity and the bulk Lorentz factor pointed out recently by \cite{Ito2021, Ito2024} and found to be in good agreement with observations. Similarly, the relative importance of polarization effects could be investigated as well. 

\section*{Acknowledgements}
Amina Trabelsi is very grateful and expresses her deep thanks to C. Lundman and A. Peer for their kind help and valuable insights. She is also very indebted to the GBM team for having provided her with the RMFIT software.    

\begin{appendices}
\section{Code diagram} \label{secA1}
\begin{center}
\includegraphics[scale=0.9]{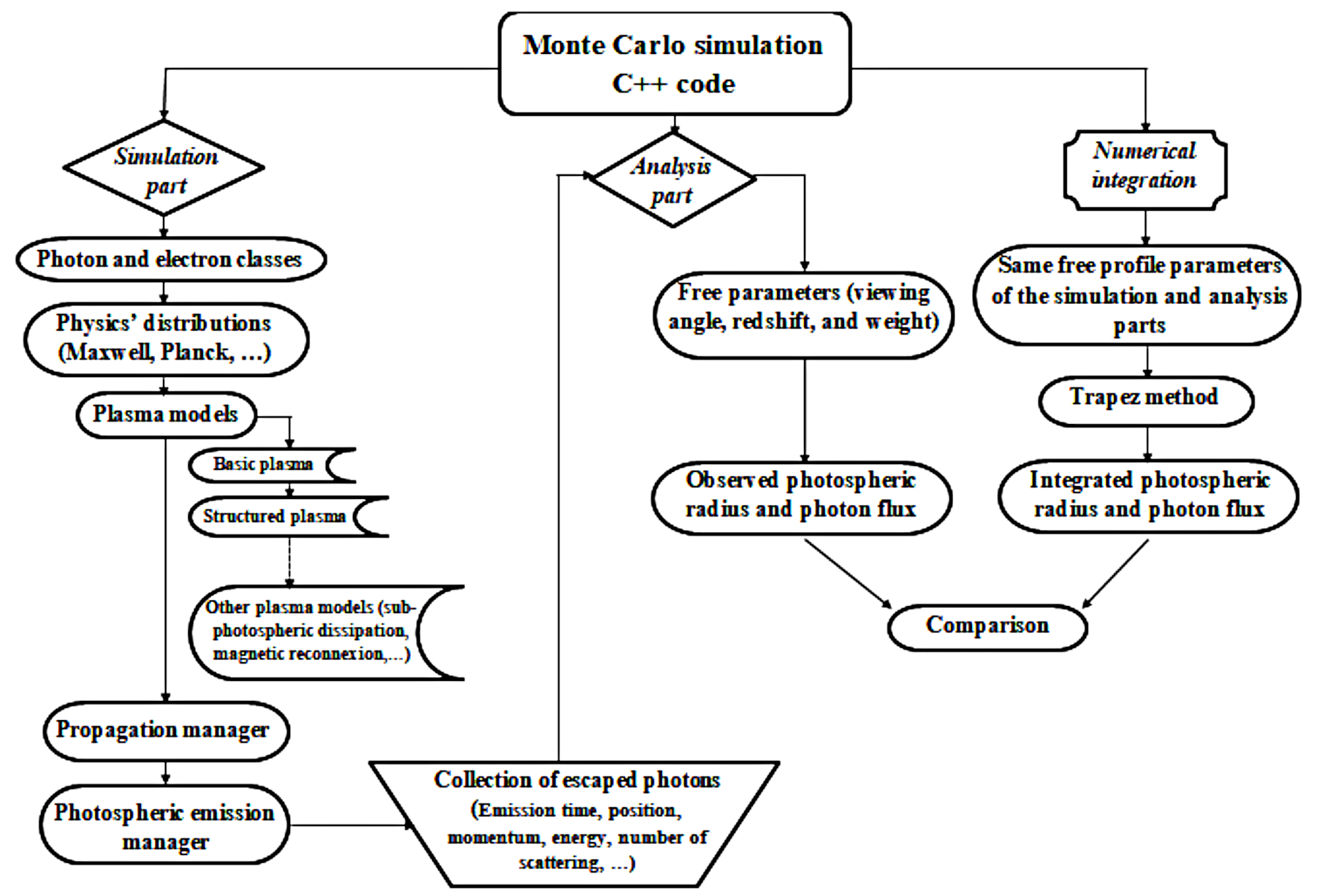}\label{diagramme}\\
Fig A : diagram of our C++ code for the M C simulation of the photospheric emission with the numerical integration code being incorporated.
\end{center}
\end{appendices}

\bibliography{bibliography}

\end{document}